\documentclass[1p,number,sort&compress]{elsarticle}
\usepackage[utf8]{inputenc}
\usepackage{amsmath}
\usepackage[labelfont=bf]{caption}
\usepackage{subcaption}
\usepackage{multirow}
\usepackage[referable]{threeparttablex}
\usepackage{footnote}
\usepackage{pdflscape}
\usepackage[colorlinks]{hyperref}
\usepackage{lineno}

\begin{document}
\begin{frontmatter}
\title{Gaussian mixture model clustering algorithms for the analysis of high-precision mass measurements}
\author[ANL]{C. M. Weber\corref{cor1}\fnref{fn1}}\ead{webe1077@umn.edu}
\author[ANL,MB]{D. Ray}
\author[ANL,MB]{A. A. Valverde}
\author[ANL,MB]{J. A. Clark}
\author[MB]{K. S. Sharma}

\address[ANL]{Physics Division, Argonne National Laboratory, Lemont, IL 60439, USA}
\address[MB]{Department of Physics and Astronomy, University of Manitoba, Winnipeg, MB R3T 2N2, Canada}

\cortext[cor1]{Corresponding author}
\fntext[fn1]{Current Address: School of Physics and Astronomy, University of Minnesota, Minneapolis, MN 55455, USA}
\begin{abstract}

The development of the phase-imaging ion-cyclotron resonance (PI-ICR) technique for use in Penning trap mass spectrometry (PTMS) increased the speed and precision with which PTMS experiments can be carried out. In PI-ICR, data sets of the locations of individual ion hits on a detector are created showing how ions cluster together into spots according to their cyclotron frequency. Ideal data sets would consist of a single, 2D-spherical spot with no other noise, but in practice data sets typically contain multiple spots, non-spherical spots, or significant noise, all of which can make determining the locations of spot centers non-trivial. A method for assigning groups of ions to their respective spots and determining the spot centers is therefore essential for further improving precision and confidence in PI-ICR experiments. We present the class of Gaussian mixture model (GMM) clustering algorithms as an optimal solution. We show that on simulated PI-ICR data, several types of GMM clustering algorithms perform better than other clustering algorithms over a variety of typical scenarios encountered in PI-ICR. The mass spectra of $^{163}\text{Gd}$, $^{163m}\text{Gd}$, $^{162}\text{Tb}$, and $^{162m}\text{Tb}$ measured using PI-ICR at the Canadian Penning trap mass spectrometer were checked using GMMs, producing results that were in close agreement with the previously published values. \par
\end{abstract}
\end{frontmatter}

\section{Introduction}\label{Intro} \par
\begin{figure}[t]
    \centering
    \includegraphics[scale=0.5]{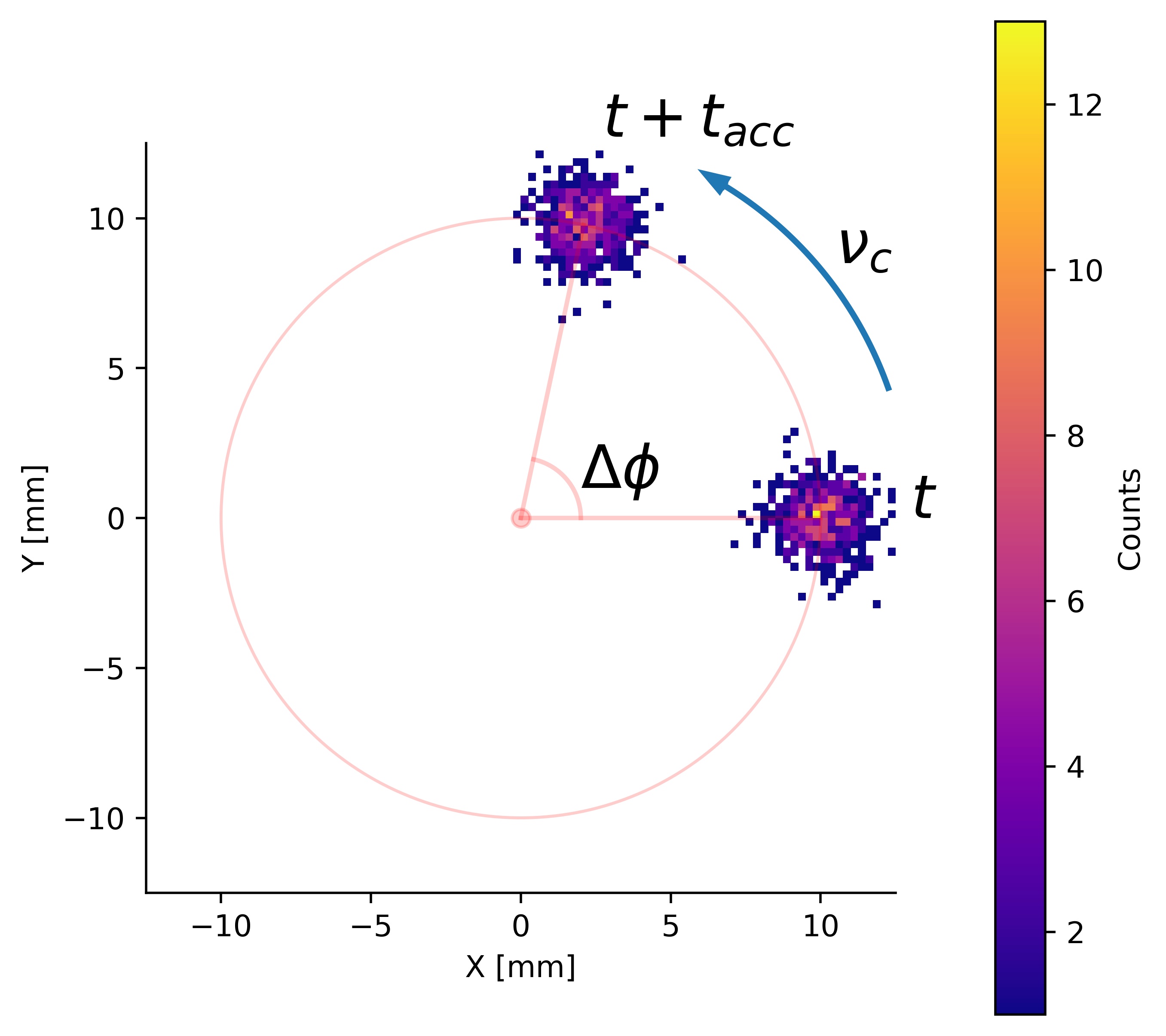}
    \caption{An illustration of a PI-ICR measurement with the initial and final spots superimposed on one spectrum. An ion species' cyclotron frequency is measured with Eq. \ref{eqn:PIICR} by determining its excess phase accumulated during some time $t_\text{acc}$. The cyclotron frequency is then used to deduce the species' mass.}
    \label{fig:PI-ICR}
\end{figure}

Penning trap mass spectrometry (PTMS) provides increasingly high-precision nuclear mass data for studies in nuclear structure, nuclear astrophysics, and fundamental physics \cite{Dilling2018}. The development of the phase-imaging ion-cyclotron-resonance (PI-ICR) technique at the SHIPTRAP facility \cite{Eitel2009, Eliseev2013, Eliseev2014} has further improved PTMS experiments, enabling mass measurements of short-lived nuclei to a precision of up to $\delta m / m = 1.4\times10^{-9}$ \cite{Karthein2019}. PI-ICR involves the measurement of an ion's cyclotron frequency $\nu_{c}$ in a trap using the relation \par
\begin{gather}
\nu_{\text{c}}=\frac{2\pi N + \Delta\phi}{2\pi t_{\text{acc}}},\label{eqn:PIICR}
\end{gather}
where $t_{\text{acc}}$ is the duration over which the ion completes $N$ complete revolutions and accumulates an excess phase of $\Delta\phi$ (Fig. \ref{fig:PI-ICR}). Because $t_{\text{acc}}$ is known and \emph{N} can be inferred, the ion's mass can be deduced by measuring only its excess phase. In these experiments, the ion's location is found by projection onto a position-sensitive microchannel plate (PS-MCP) detector \cite{Eitel2009, Eliseev2013, Eliseev2014, Nesterenko2018, OrfordNIMB2020, Karthein2021}. For details of PI-ICR's implementation at the Canadian Penning trap mass spectrometer (CPT) at Argonne National Laboratory's ATLAS accelerator facility, see Ref. \cite{OrfordNIMB2020}. \par

The data analysis procedure in a PI-ICR experiment begins with detecting individual ion hits on the PS-MCP. Because the locations on the detector are linearly related to locations in the trap, data sets are representations of the locations of ions in the trap (Fig. \ref{fig:PI-ICR}). Next, manual data cuts may be made to highlight specific temporal and spatial regions of the data and obtain a clean signal \cite{Orford2019, Karthein2020, Karthein2021}. Finally, because ions of the same cyclotron frequency appear as ``clusters" in the data set, the resulting data is fit to a model to determine the centers of the clusters, which can be translated to the necessary phase measurements for ion species.

An ideal data set would consist of a single, normally-distributed, spherical (meaning it has the same standard deviation in the \emph{x-} and \emph{y-}dimensions) cluster with no noise around it. In this case, determining the cluster center would be nearly trivial, and any number of methods could be used to determine the cluster center such as fitting a multivariate Gaussian curve or two one-dimensional Gaussian curves to the data. Indeed, both of these methods have been used successfully in PI-ICR experiments \cite{Eliseev2013, OrfordNIMB2020, Karthein2020}, although it has been shown that for simulated data the multivariate Gaussian model reduces uncertainty in the ions' phase determination relative to using two one-dimensional Gaussians \cite{Karthein2021}. The authors of Ref. \cite{Karthein2021} also showed with simulated data that fitting a univariate Gaussian model to only the phase dimension of the ion positions in polar coordinates reduces the uncertainty by up to a factor of 10 relative to the multivariate Gaussian fit. 

\begin{figure*}[t]
     \centering
     \begin{subfigure}[t]{0.3\textwidth}
         \centering
         \includegraphics[width=\textwidth]{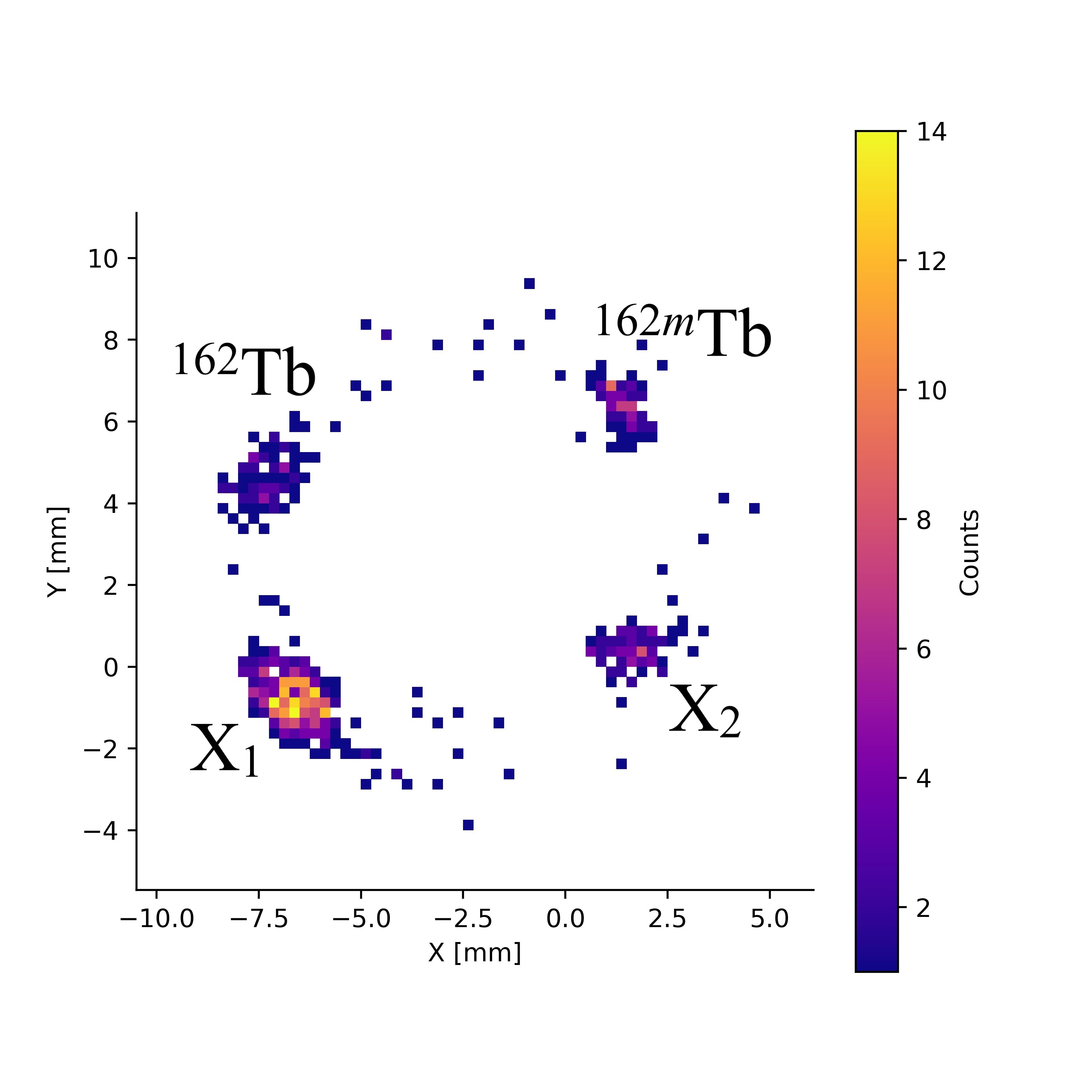}
         \caption{The algorithm begins with an unlabelled data set. Pictured is a spectrum from an experiment conducted with the CPT measuring the ground and isomeric masses of $^{162}$Tb as described in Ref. \cite{OrfordPRC2020}. $\text{X}_1$ and $\text{X}_2$ are isobaric contaminants in the beam.}
         \label{fig:GMM a}
     \end{subfigure}
     \hfill
     \begin{subfigure}[t]{0.35\textwidth}
         \captionsetup{width=0.8\textwidth}
         \centering
         \includegraphics[width=\textwidth]{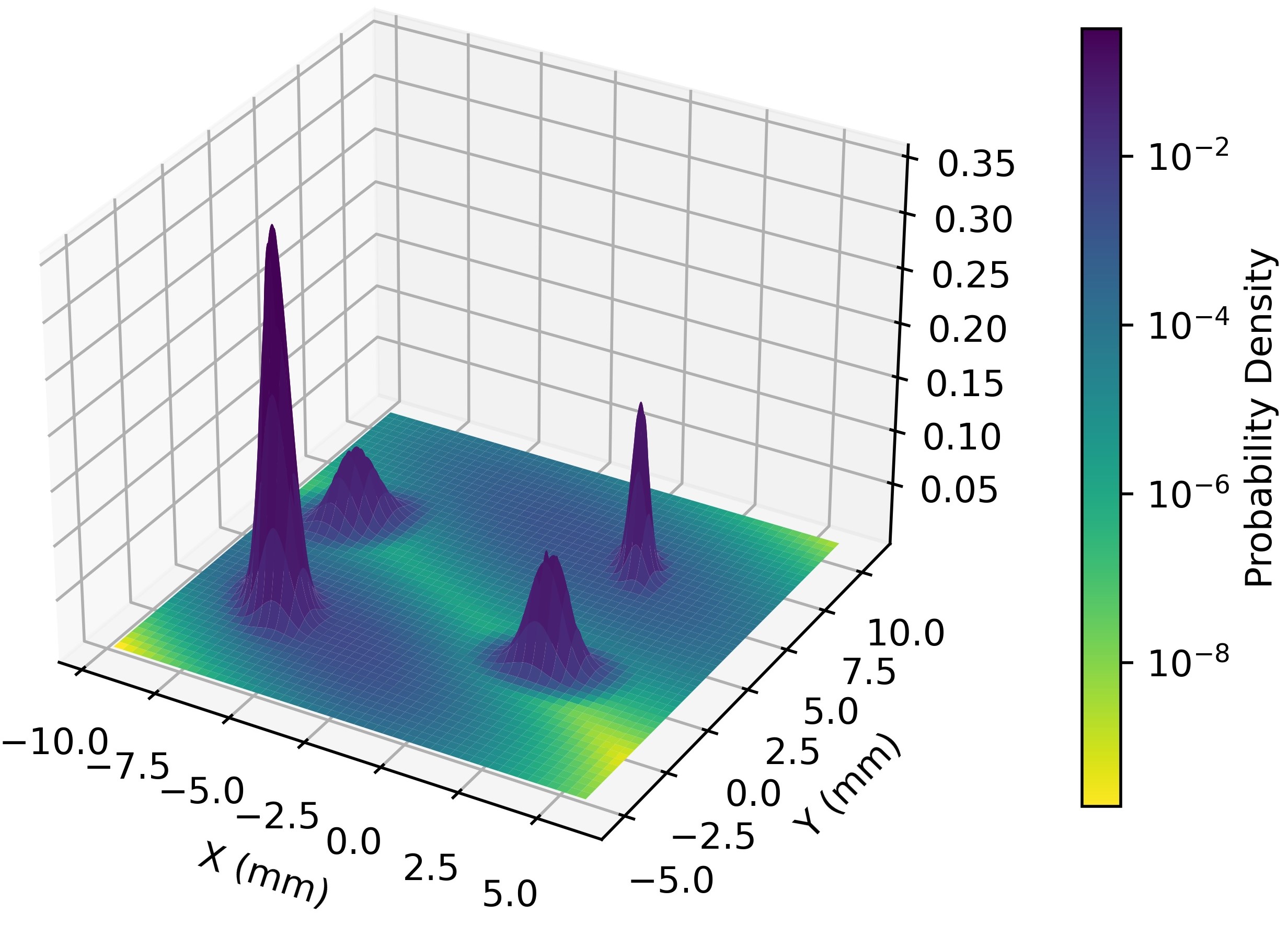}
         \caption{For several test values of \emph{K}-components, a GMM is fit to the data. \emph{K} refers to the number of multivariate Gaussian components in the model.}
         \label{fig:GMM b}
     \end{subfigure}
     \hfill
     \begin{subfigure}[t]{0.3\textwidth}
         \centering
         \includegraphics[width=\textwidth]{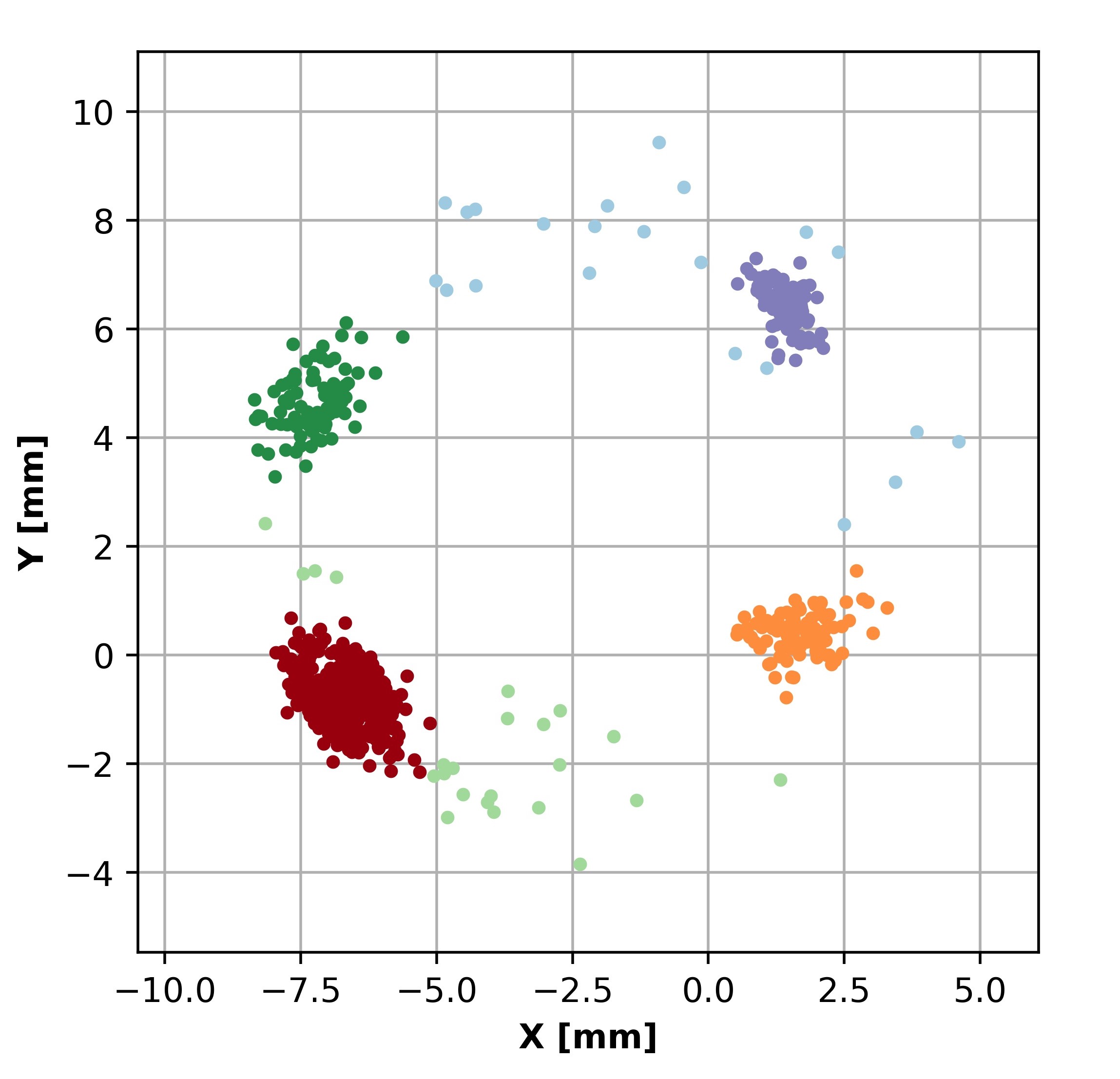}
         \caption{After using the BIC to determine an optimal value for \emph{K}, the data is assigned to clusters based on the Gaussian component with respect to which it has a maximum log-likelihood of occurring. This shows a successful clustering result because two ``noise" clusters (the lightest colored spots) are identified and separated from the four data clusters.}
         \label{fig:GMM c}
     \end{subfigure}
     \caption{An example of how maximum likelihood estimation is used to fit a GMM to PI-ICR data.}
     \label{fig:GMM Algorithm}
\end{figure*}

However, many PI-ICR data sets contain multiple spots, non-spherical spots, or noise, so a single multivariate Gaussian model is not complex enough to capture the data. Consider, for example, the empirical data set shown in Fig. \ref{fig:GMM a}, in which there are four non-spherical spots with a bit of noise as well. Even if the data set was restricted spatially to focus on a single cluster, systematic uncertainties could be introduced by the subjective assignment of boundaries to the subset. These systematic uncertainties could be especially prevalent if certain challenging but oft-encountered situations are present, such as if there are multiple or overlapping clusters in the subset, if the subset is non-spherical, or if there is significant noise remaining in the subset. The need for a method of systematically partitioning a data set into subsets based on a common characteristic, or \emph{clustering}, is therefore critical for improving the precision of PI-ICR experiments.

Unsupervised machine learning algorithms that ``cluster" data automatically, known as \emph{clustering algorithms}, have thus far received little attention with regard to mass spectrometry despite their being possible solutions to the problem outlined above. Clustering algorithms are a type of unsupervised machine learning algorithm that divide data sets into clusters such that data points that are more similar based on some metric, like separation or density, are clustered together, and dissimilar data points are either clustered separately or omitted from all clusters. Because one of the challenges in PI-ICR data analysis is separating data sets into clusters based on ion species, and ion species cluster together spatially in the data, clustering algorithms can therefore be used to automatically divide data sets into clusters, avoiding the introduction of subjective data cuts. In this paper, we present Gaussian mixture models as candidate clustering algorithms for PI-ICR data, and we demonstrate the results of testing them on both simulated and empirical data.

\section{Gaussian mixture models} \label{background}
A clustering algorithm for PI-ICR experiments should satisfy several criteria. It must function with spatial data, and do well with non-spherical clusters. Density-based clustering algorithms, such as DBSCAN and Mean Shift, as well as their variants \cite{Lloyd1982, Ester1996, Hinneburg1998, Dash2001, Comanciu2002}, fit both of these requirements. In general they work by identifying the peak densities in a data set, and then building clusters out from the peaks, stopping when the density of points at the cluster boundary falls below a density threshold parameter. Yet it is this density threshold parameter which determines the cluster boundaries that makes these algorithms not optimal for PI-ICR data, because it requires an assumption to be made about cluster size, that is, where a cluster boundary should be.

Another popular clustering algorithm, \emph{K}-Means, avoids the problems of density-based clustering algorithms by only requiring one parameter: the number of clusters \emph{K} to find \cite{Lloyd1982}. This algorithm begins by randomly selecting \emph{K} cluster centers, and then clustering data points to their closest centers. Next, the mean value of each cluster is calculated and the centers are moved to these locations. The two steps of data assignment and center updating repeats until convergence. The main drawback of the \emph{K}-Means algorithm, as well as most of its variants, is that it still requires an initial guess to be made about the number of clusters in the data set, and there is no consensus about how best to solve this problem \cite{Bezdek1984, Pal1995, Wu2009}. Furthermore, \emph{K}-Means is known to perform poorly with non-spherical clusters, which are the norm in PI-ICR \cite{Wu2009}.

With Gaussian mixture models as with \emph{K}-Means, the number of user-defined input parameters can be reduced to one, namely, the number of components to use \emph{K}. The difference between Gaussian mixture models and \emph{K}-Means, however, is that with Gaussian mixture models, there are several reputable methods for determining \emph{K} and for determining the other model parameters. Here we elaborate on the two methods we tested on PI-ICR data: maximum-likelihood (ML) estimation, which uses the Ex\-pectation-Maximization (EM) algorithm, and Variational Bayesian Inference, both of which we implemented using the Python package \emph{scikit-learn} \cite{scikit-learn}. Each algorithm was broken down additionally into two flavors corresponding to the two coordinate systems, Cartesian and polar, that are most convenient to use for PI-ICR data.

To estimate the parameters for a GMM with a given number of \emph{K} components, we used the EM algorithm, which consists of two steps that repeat until convergence following parameter initialization:
\begin{enumerate}
    \item Expectation: given the current estimate of parameters, calculate the expected value of the log-likelihood of the data samples to assign each sample to a component of the GMM. 
    \item Maximization: given the expected log-likelihood values and cluster assignments, update the parameters to maximize the complete-data log-likelihood, giving them the value that maximizes the likelihood that the data has the expected labels \cite{Reynolds2009, Millar2011}.
\end{enumerate}
To determine \emph{K}, we fit GMMs to several test values of \emph{K} and selected the one that minimized the Bayesian Information Criterion (BIC), signifying that it is the most likely model for the data \cite{Schwarz1978, Stine2004, Dziak2020}. Finally, we assigned each data sample to a cluster based on the Gaussian component with respect to which it had a maximum log-likelihood of occurring, with cluster centers and uncertainties given by the model parameters. A visualization of this procedure is shown in Fig. \ref{fig:GMM Algorithm}.

For the Variational Bayesian Inference method, we follow the procedures as outlined in Refs. \cite{Bishop2006, Blei2006} by first assigning \emph{a priori} distributions to the parameters of our Bayesian Gaussian mixture (BGM) model. We used the stick-breaking representation of the Dirichlet process, a Gaussian distribution, and the Wishart distribution as priors for our component weights, means, and precisions, respectively. Next, we again used the EM algorithm to fit our model, except that in the maximization step, we set the model parameters to their expectation values calculated from their respective posterior distributions, which are updated according to the likelihoods generated in the expectation step. Because this approach allows for the component weights to approach 0 and thus for there to be components to which no samples belong, we initialized our model with an arbitrarily high number of components and determined \emph{K}-components to be the number of components to which there belongs at least one sample. Data samples, centers, and uncertainties were assigned in the same way as with the GMM methods.

The fifth method we tested, the so-called ``Phase-First Gaussian Mixture", is a version of GMM in which we consider that for PI-ICR analysis, we are primarily concerned with the phase dimension of the ions. First, we fit GMMs to the phase dimension of the data, again using the BIC to find the number of components to use. We then used the fit centers and sample assignments as the initialization for a GMM fit to the full polar data set. In the second fit, we fixed the phase dimension of the centers to be the phases given by the phase-only fit, and then determined the cluster assignments, centers, and uncertainties as with the other models.

\section{GMM evaluation on simulated PI-ICR data} \label{Simulated Data}

\begin{figure*}[t!]
     \centering
     \begin{subfigure}[t]{0.3\textwidth}
         \centering
         \includegraphics[width=\textwidth]{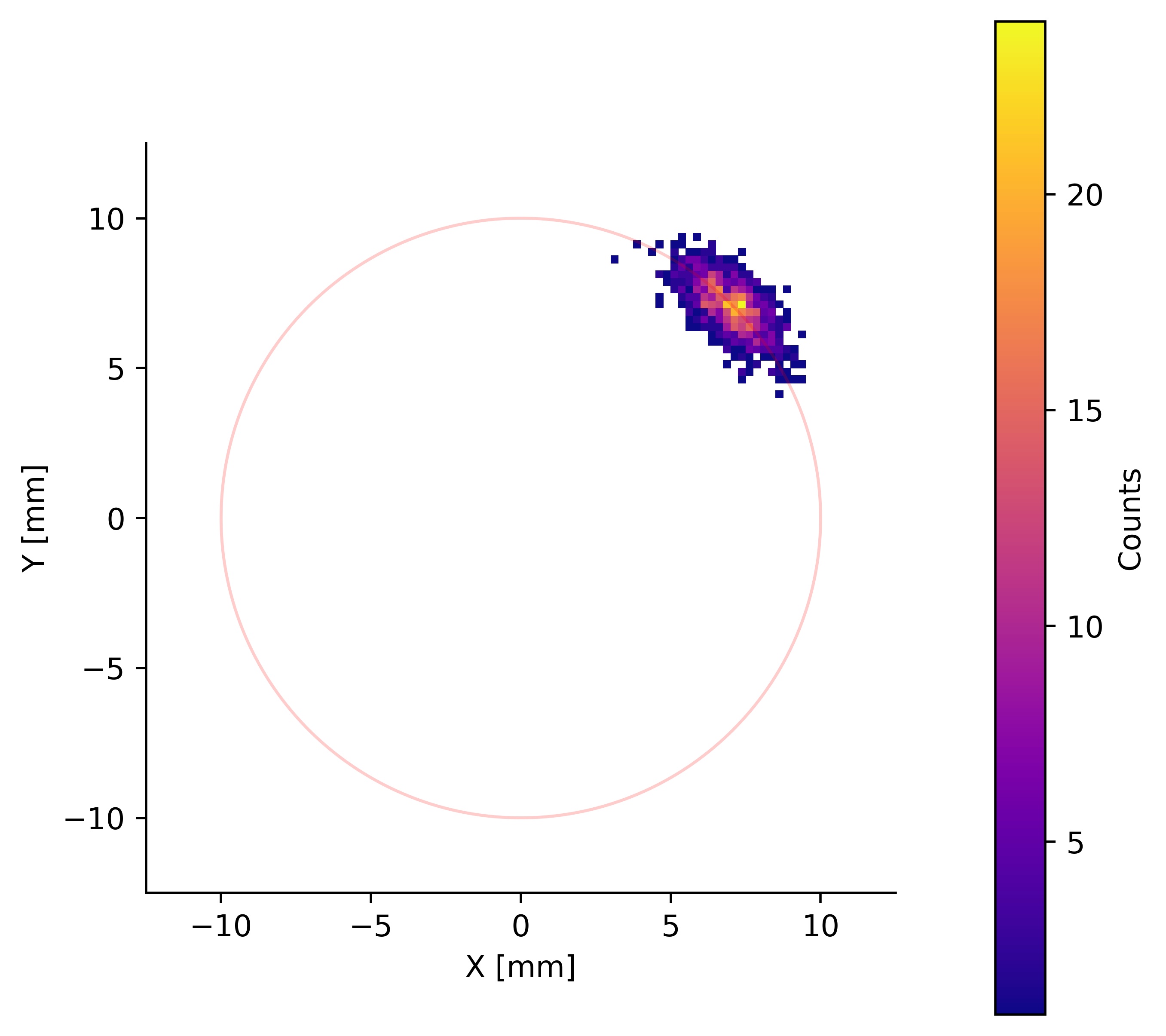}
         \caption{A simulated scenario with a single spot composed of 1000 samples drawn from polar coordinates centered at a phase of $45^{\circ}$ with ellipticity 1. Radius and spot size were chosen to best replicate experimental data.}
         \label{fig:single spot}
         \quad
     \end{subfigure}
     \hfill
     \begin{subfigure}[t]{0.3\textwidth}
         \centering
         \includegraphics[width=\textwidth]{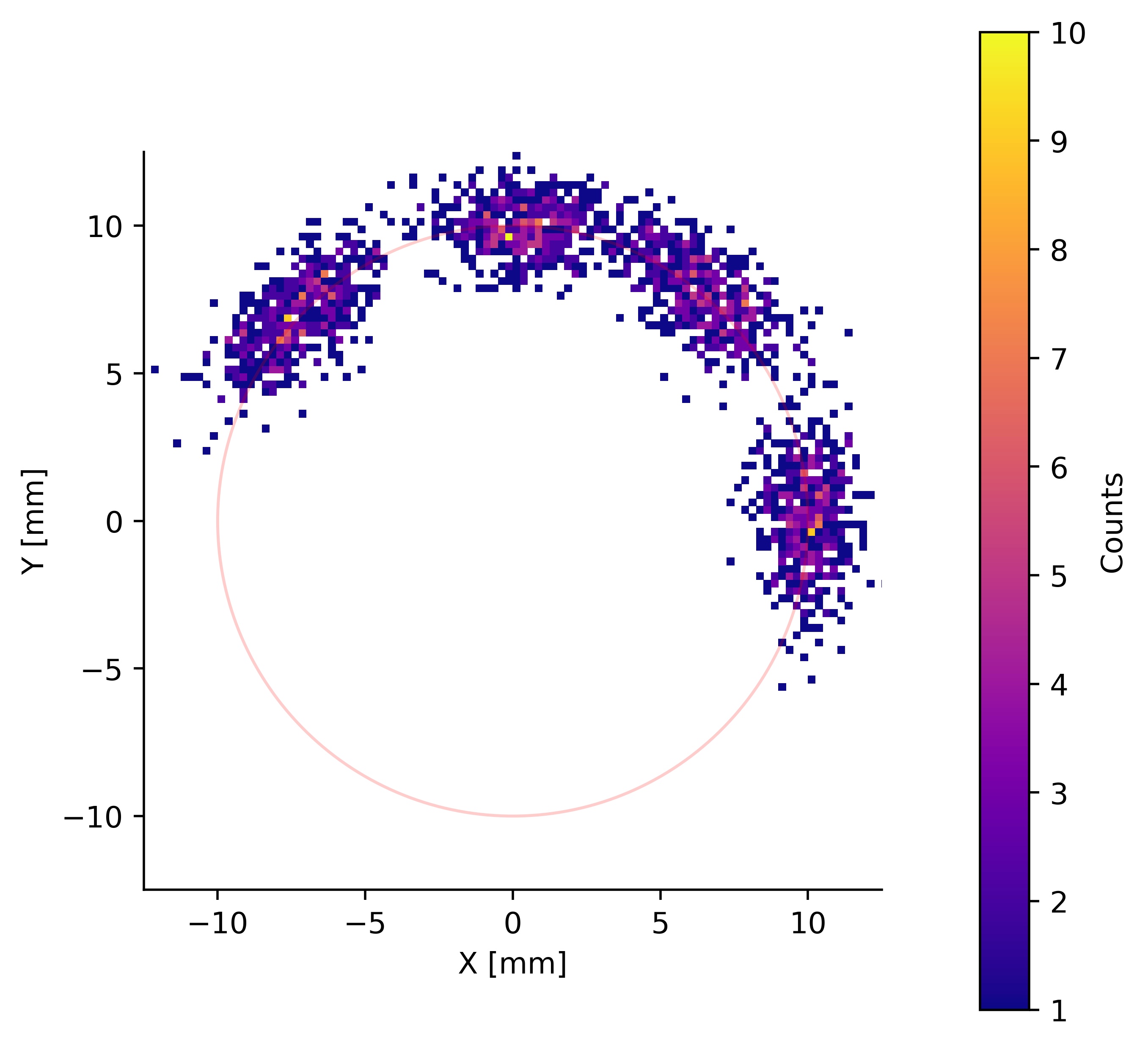}
         \caption{A scenario with multiple spots of 500 ions each drawn from Cartesian coordinates, separated by $5\sigma$, and with ellipticity 2.}
         \label{fig:multispot}
         \quad
     \end{subfigure}
     \hfill
     \begin{subfigure}[t]{0.3\textwidth}
         \centering
         \includegraphics[width=\textwidth]{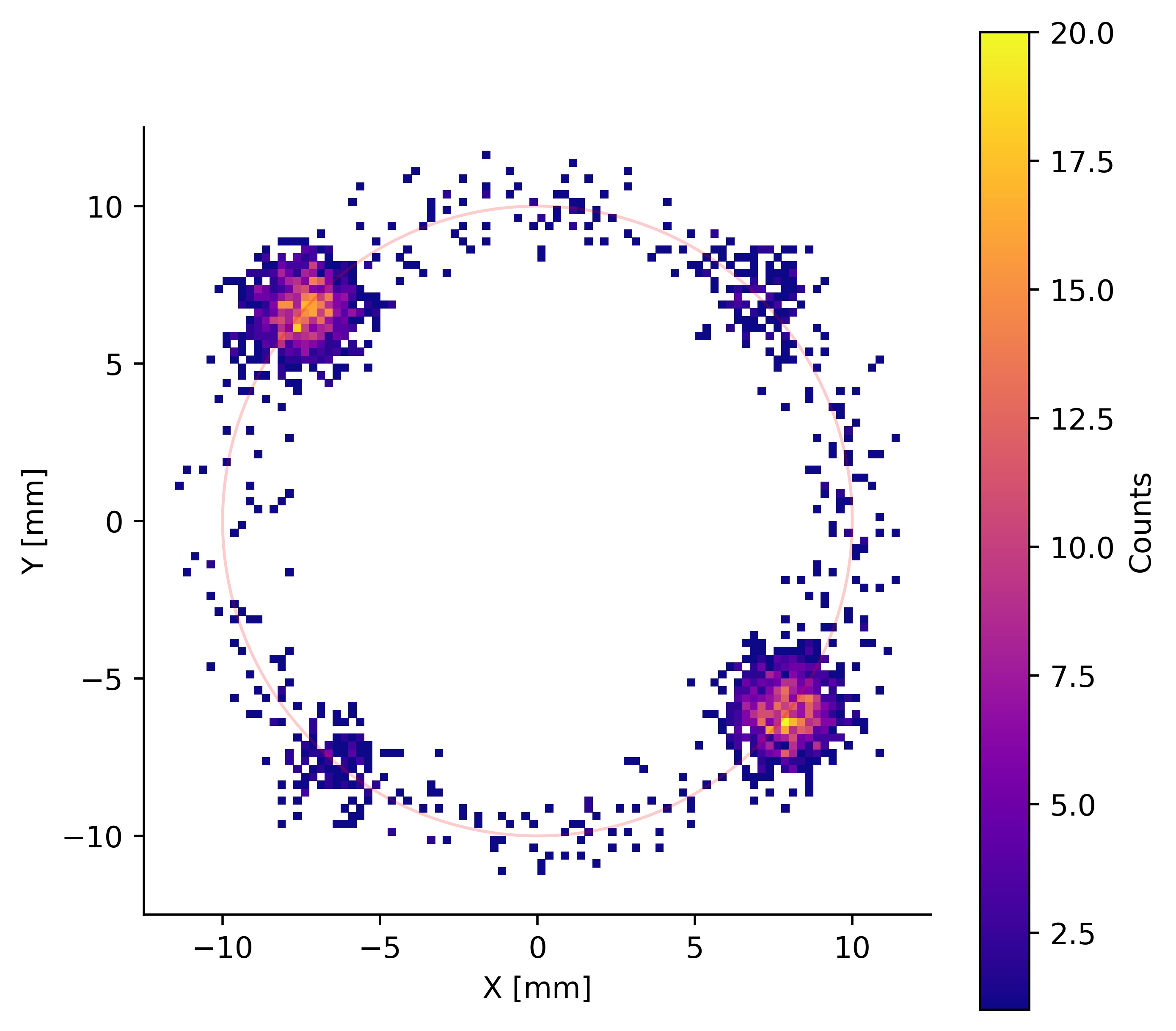}
         \caption{A scenario with noise on the ring in which the ratio of noise spots to the number of spots in the smallest cluster is 1. Spot ellipticities are all 1, and the least dense spots are composed of 100 samples, whereas the more dense spots are composed of 10 times that.}
         \label{fig:noisy}
     \end{subfigure}
     \caption{Examples of simulated PI-ICR data.}
     \label{Examples of spectra}
\end{figure*}

Our initial tests of the Gaussian mixture models were on simulated PI-ICR data in Monte Carlo simulations consisting of 1000 repetitions. The simulated data was generated to most accurately approximate experimental data. This consideration led us to set our ring radius and spot variance (and thus apparent size) accordingly. To generate the individual ion locations, we defined three broad subsets of PI-ICR data, as well as variables within each subset to test the models against. The first subset of data consisted of spectra in which there is only one spot and no noise (Fig. \ref{fig:single spot}). Within the single-spot subset, we tested data sets where the spot had $10^2,$ $10^3,$ and $10^4$ samples. Samples were drawn from polar or Cartesian coordinates, and spots had an ellipticity of 1 or 2. The spot centers were at phases of $0^{\circ}-45^{\circ}$ in $15^{\circ}$ increments, measured counter-clockwise from the horizontal line $x=0$. We also tested spots with additional rotations applied relative to the tangent to the ring at the location of the spot center at values of $0^{\circ}-45^{\circ}$ in $15^{\circ}$ increments. Additional phase locations and rotations were not tested due to the rotational symmetries of the algorithms.

\begin{figure*}[t]
     \centering
     \begin{subfigure}[t]{0.3\textwidth}
         \centering
         \includegraphics[width=\textwidth]{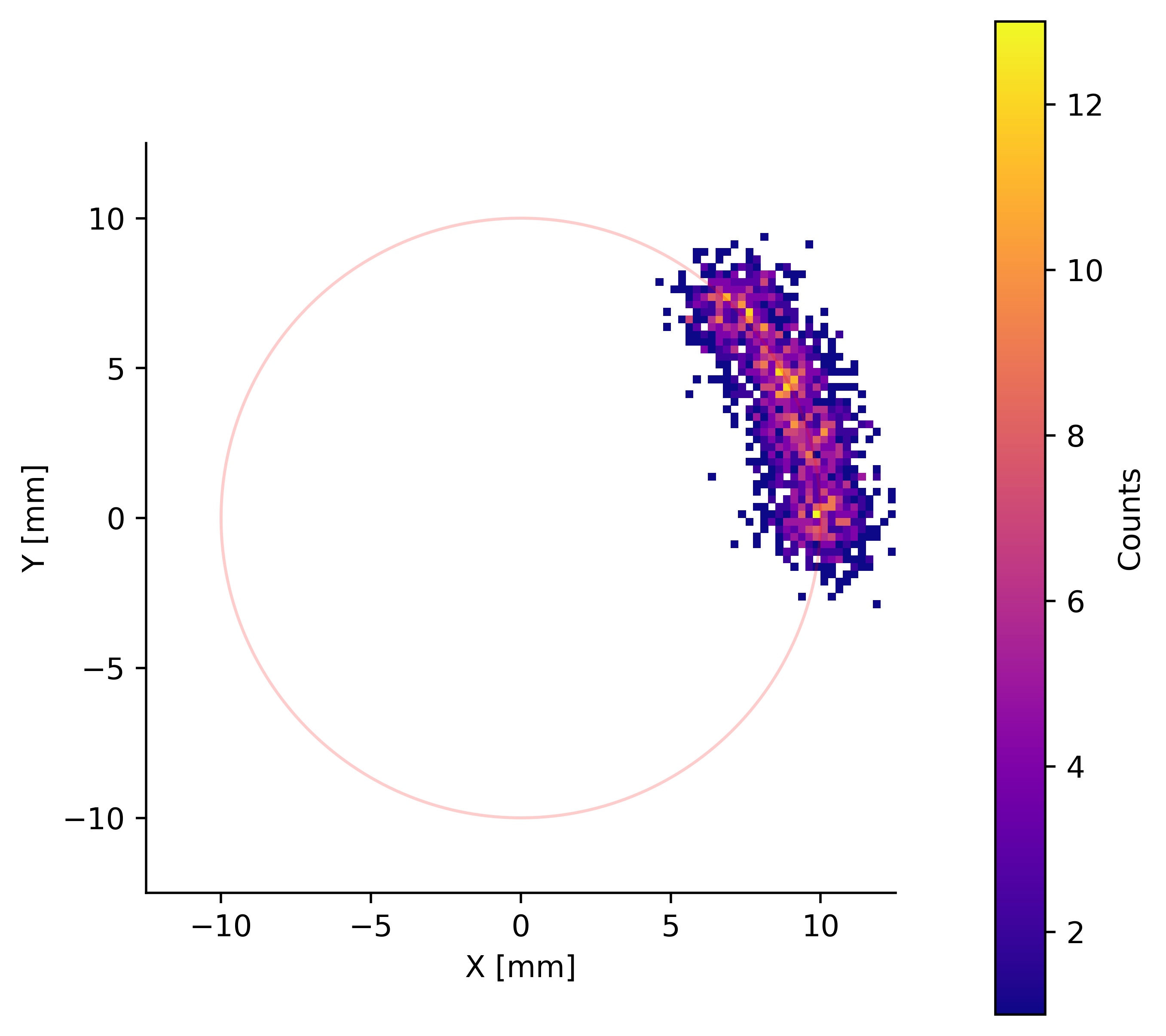}
         \caption{3$\sigma$}
         \label{fig:3 sigma}
     \end{subfigure}
     \hfill
     \begin{subfigure}[t]{0.3\textwidth}
         \centering
         \includegraphics[width=\textwidth]{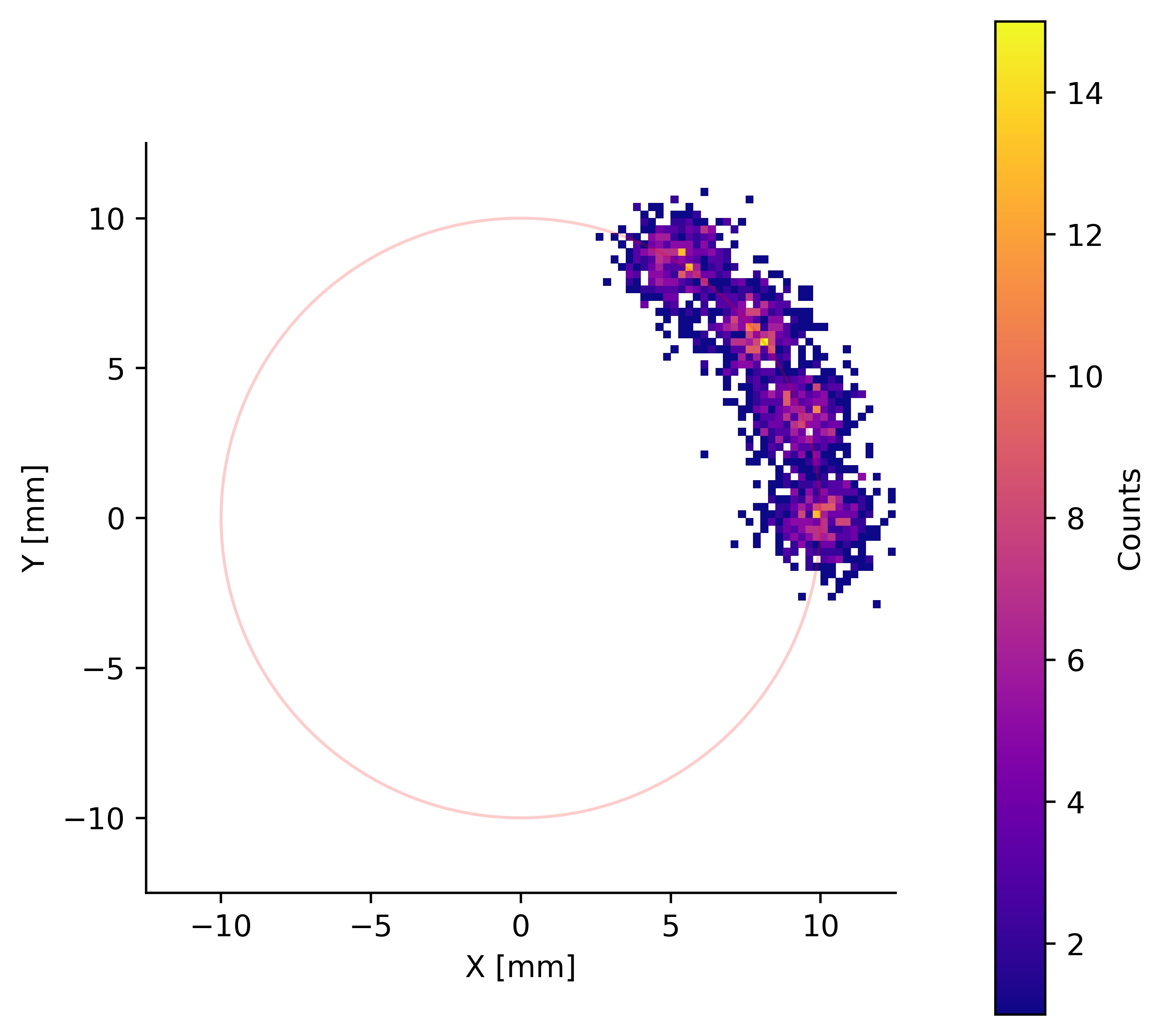}
         \caption{4$\sigma$}
         \label{fig:4 sigma}
     \end{subfigure}
     \hfill
     \begin{subfigure}[t]{0.3\textwidth}
         \centering
         \includegraphics[width=\textwidth]{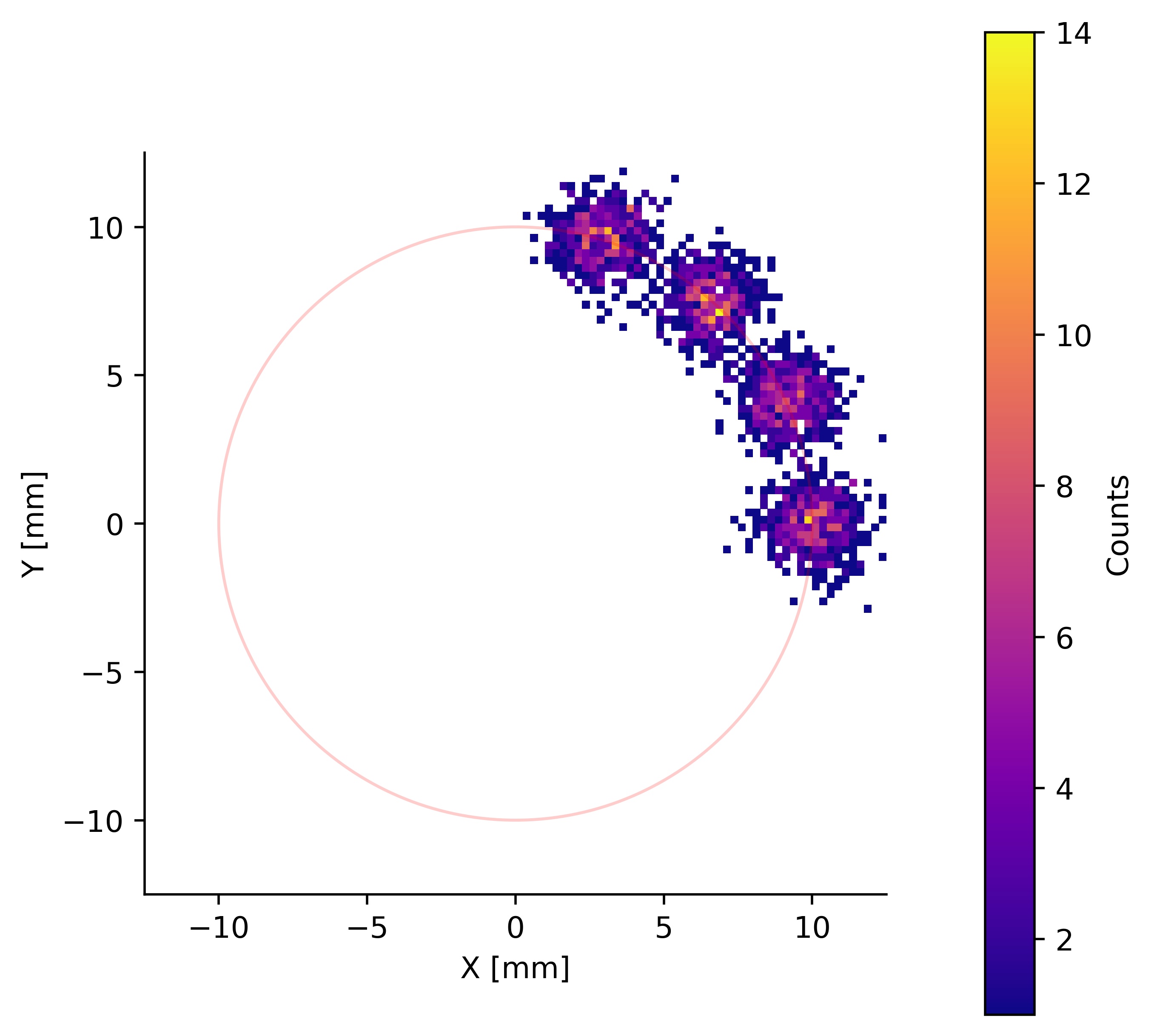}
         \caption{5$\sigma$}
         \label{fig:5 sigma}
     \end{subfigure}
     \caption{Various spot separations in terms of the standard deviation $\sigma$ in the distribution of the ions' phase coordinates.}
     \label{fig:separations}
\end{figure*}

The second subset of data consisted of spectra in which there were at least two spots, and no noise (Fig. \ref{fig:multispot}), corresponding to the experimental scenario in which there are multiple isobars in the beam. We tested scenarios in which there were 2, 4, or 8 spots. The spot separation, measured in terms of the standard deviation $\sigma$ of the phase values of the samples in each spot, was set at 3$\sigma$, 4$\sigma$, or 5$\sigma$. For all data sets, we placed the first spot on the horizontal line $x=0$ and subsequent spots on the main ring at the given phase separation counter-clockwise from the first spot (Fig. \ref{fig:separations}). We also tested scenarios in which spots had different intensities. The spot intensity variable, which describes the ratio between the number of samples in the more dense spots to the number of samples in the less dense spots, was evaluated at 1 and 10. For reference, the spot intensity of the blue cluster in Fig. \ref{fig:GMM c} relative to the other clusters is 4.4. Beginning with the first spot, all odd-numbered spots were assigned to be less dense and were given 100 samples, whereas the even-numbered spots were given a number of samples based on the spot intensity variable.

The third subset of data consisted of spectra in which there was one spot in each quadrant consisting of 100 samples drawn from Cartesian coordinates with ellipticity one and noise on the ring (Fig. \ref{fig:noisy}), corresponding to experiments in which there are multiple species in the beam and in which manual noise cuts fail to remove all noise from the data set. One of the two variables considered for these scenarios was noise intensity, which we defined to be the ratio between the number of noise samples to the number of samples in the smallest spot. For example, the noise intensity of the spectrum in Fig. \ref{fig:GMM Algorithm} is 0.55. We tested noise intensity values from 0-15 in whole numbers. The second variable we considered for scenarios with noise on the ring was spot intensity, which we defined the same as with the scenarios with multiple spots. Spot intensity values evaluated were 1 and 10.

Each combination of variables was used to generate a set of 1000 spectra to be used in the Monte Carlo simulations. Each of the five clustering methods described above was then used to cluster the data, and each method was evaluated for precision and accuracy relative to the well-known Mean Shift clustering algorithm, which was used by the authors of Ref. \cite{OrfordNIMB2020} in their PI-ICR data analysis. When using the Mean Shift algorithm, spot centers and uncertainties were calculated by fitting univariate Gaussian curves to the \emph{x} and \emph{y} dimensions of the samples in a cluster, because the Mean Shift algorithm merely clusters data and doesn't determine these values.

\begin{figure*}[ht!]
     \centering
     \begin{subfigure}[t]{0.3\textwidth}
         \centering
         \includegraphics[width=\textwidth]{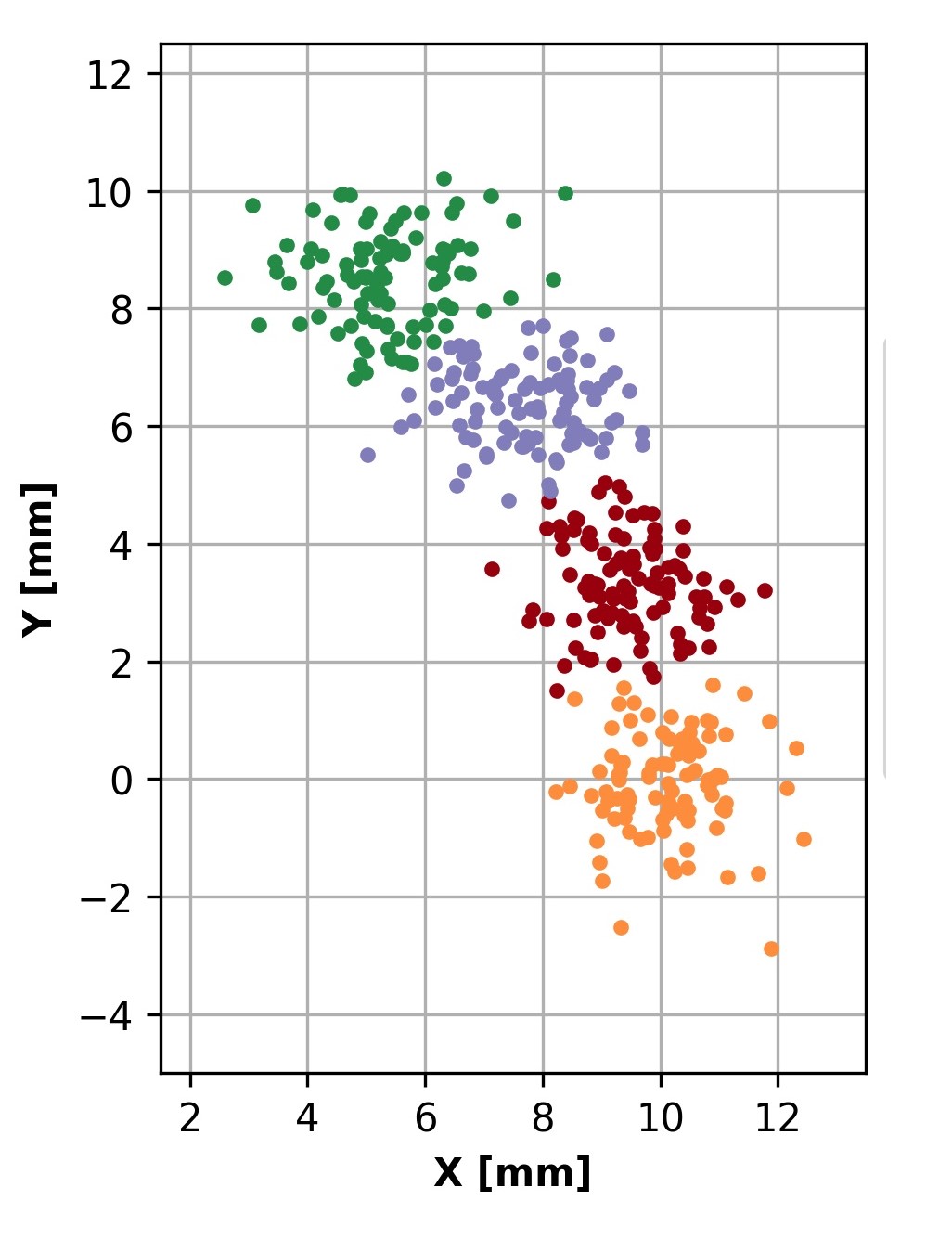}
         \caption{EM algorithm with data in Cartesian coordinates}
         \label{fig:GMMCart compare}
         \quad
     \end{subfigure}
     \hfill
     \begin{subfigure}[t]{0.3\textwidth}
         \centering
         \includegraphics[width=\textwidth]{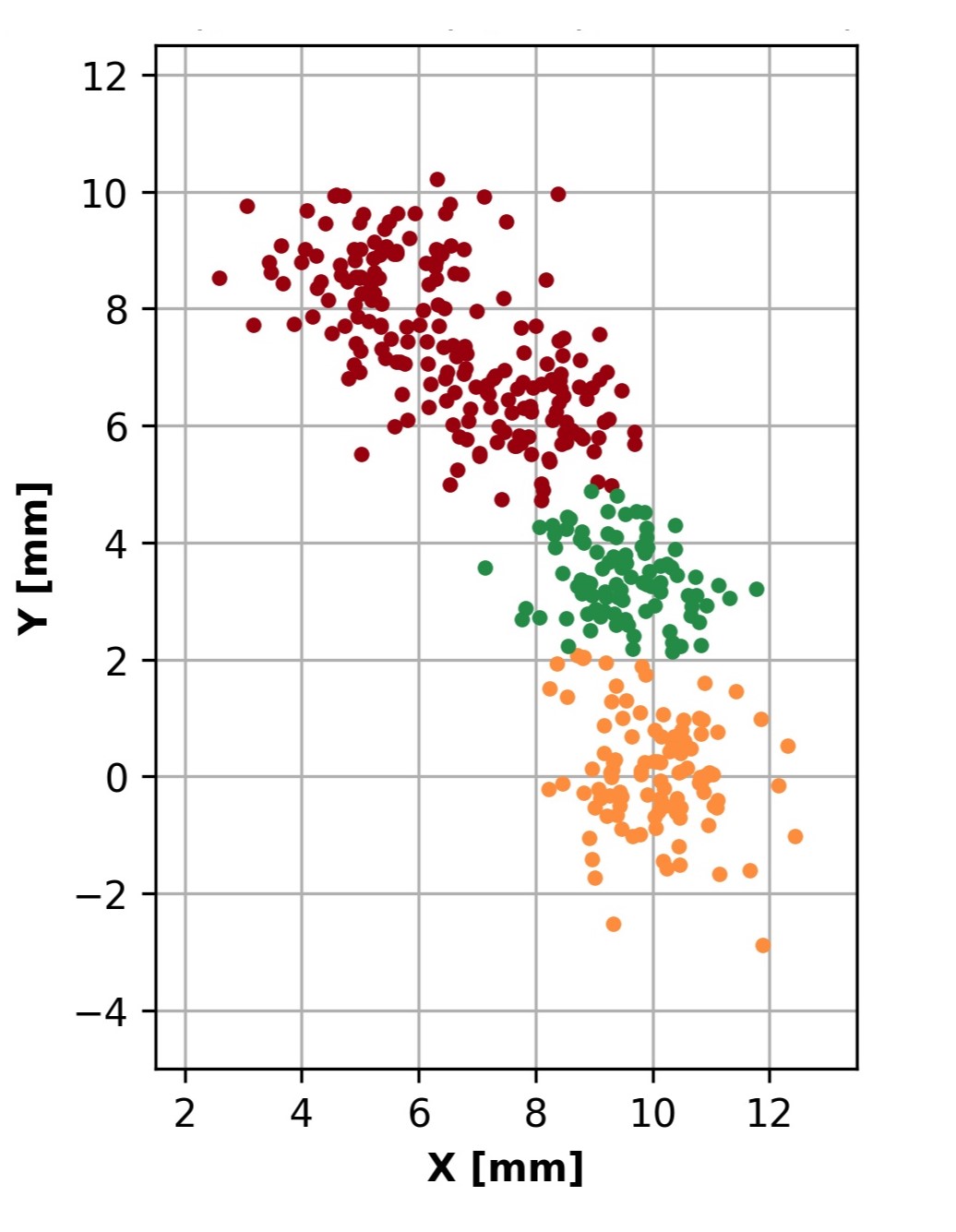}
         \caption{Variational Bayesian algorithm with data in Cartesian coordinates}
         \label{fig:BGMCart}
     \end{subfigure}
     \hfill
     \begin{subfigure}[t]{0.3\textwidth}
         \centering
         \includegraphics[width=\textwidth]{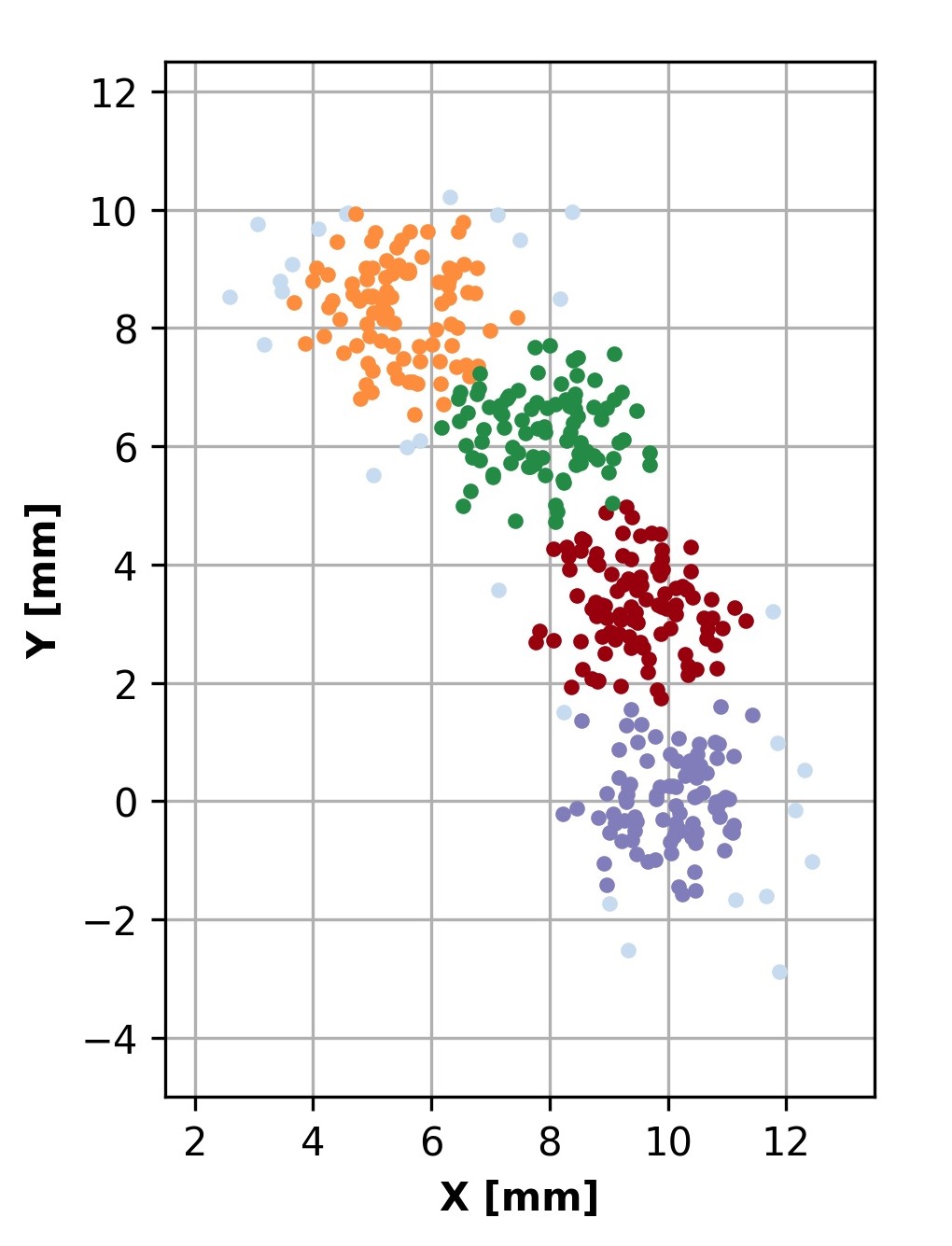}
         \caption{Mean Shift algorithm}
         \label{fig:MeanShift}
     \end{subfigure}\\
     \begin{subfigure}[t]{0.3\textwidth}
         \centering
         \includegraphics[width=\textwidth]{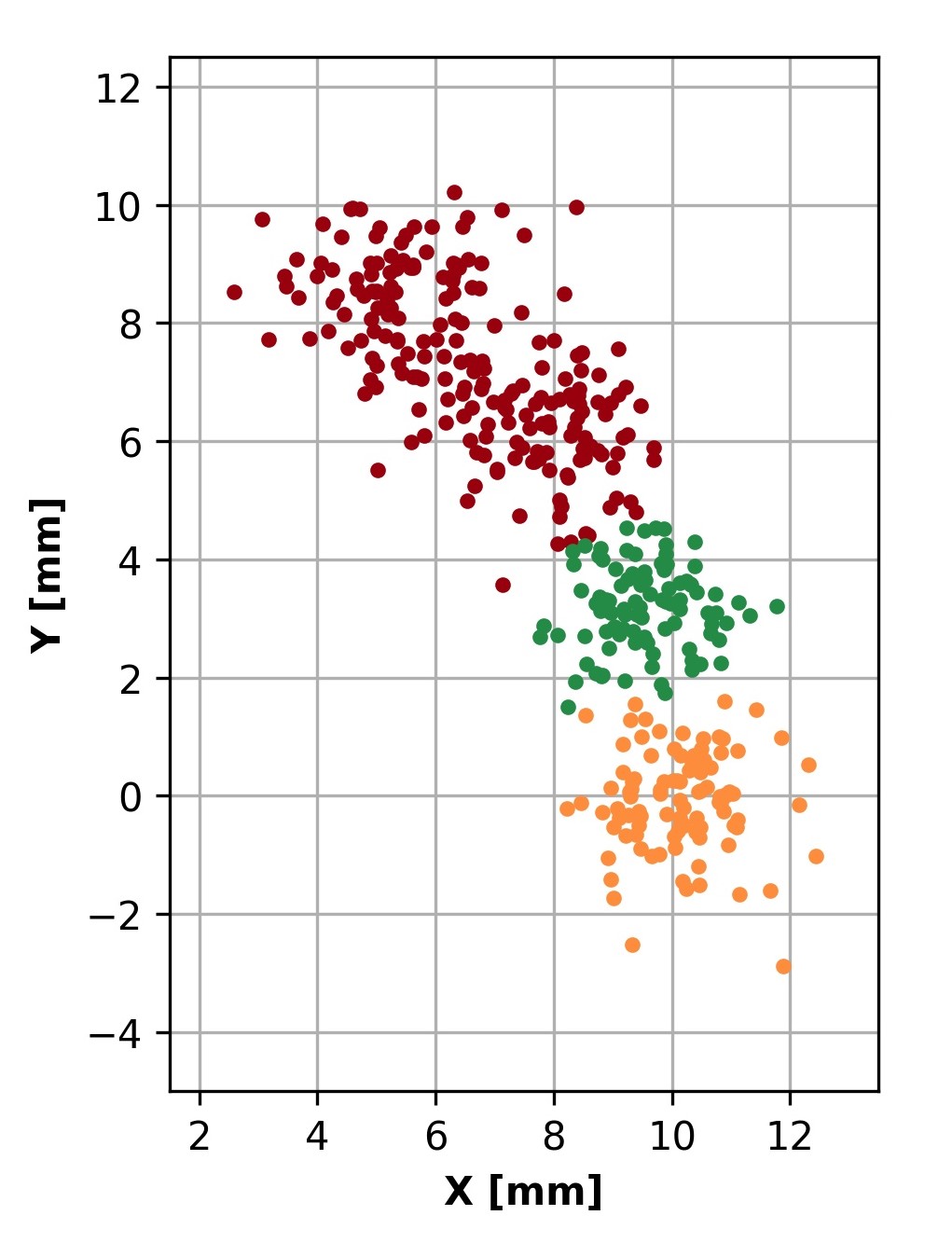}
         \caption{EM algorithm with data in polar coordinates}
         \label{fig:GMMPolar}
         \quad
     \end{subfigure}
     \hfill
     \begin{subfigure}[t]{0.3\textwidth}
         \centering
         \includegraphics[width=\textwidth]{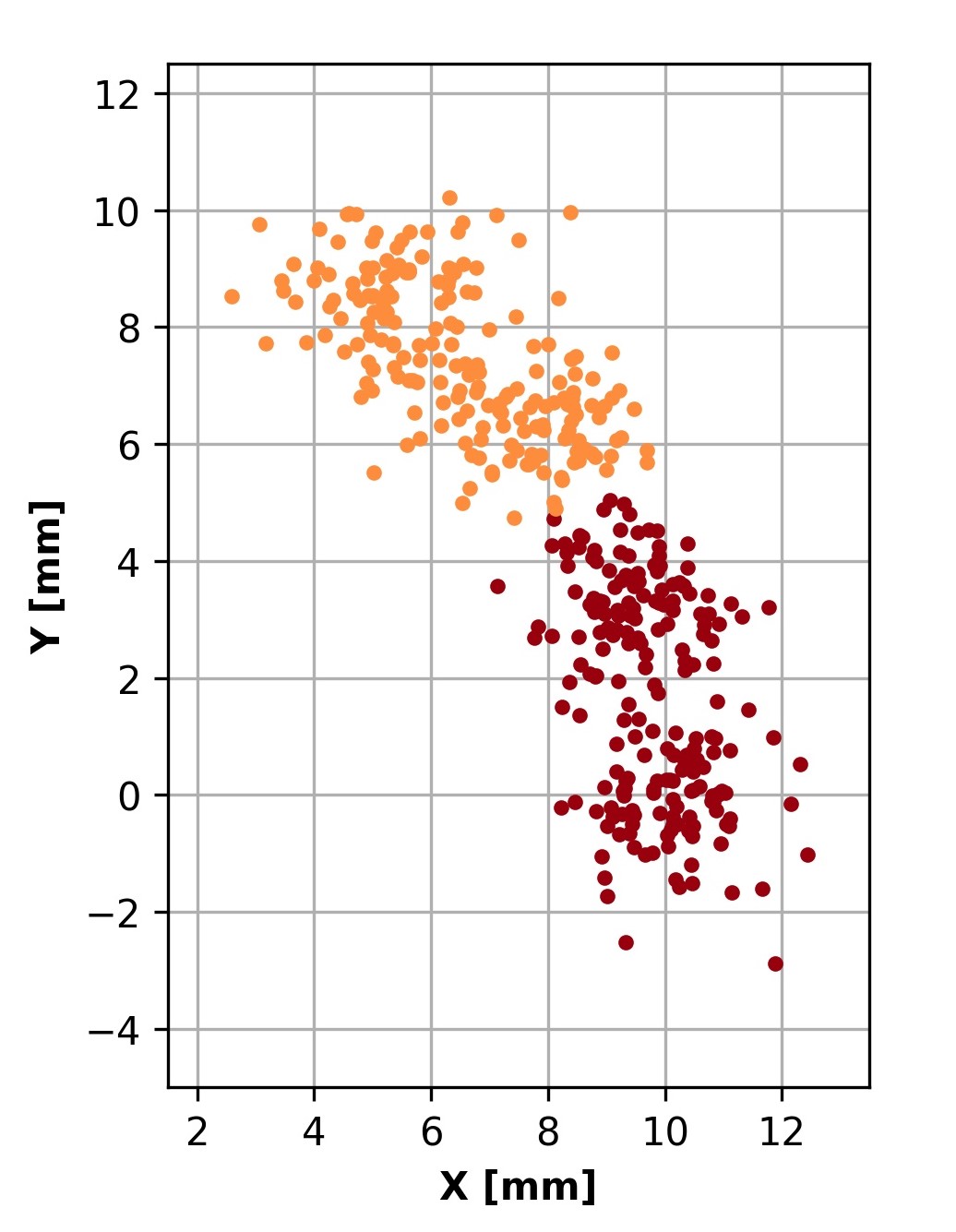}
         \caption{Variational Bayesian algorithm with data in polar coordinates}
         \label{fig:BGMPolar}
     \end{subfigure}
     \hfill
     \begin{subfigure}[t]{0.3\textwidth}
         \centering
         \includegraphics[width=\textwidth]{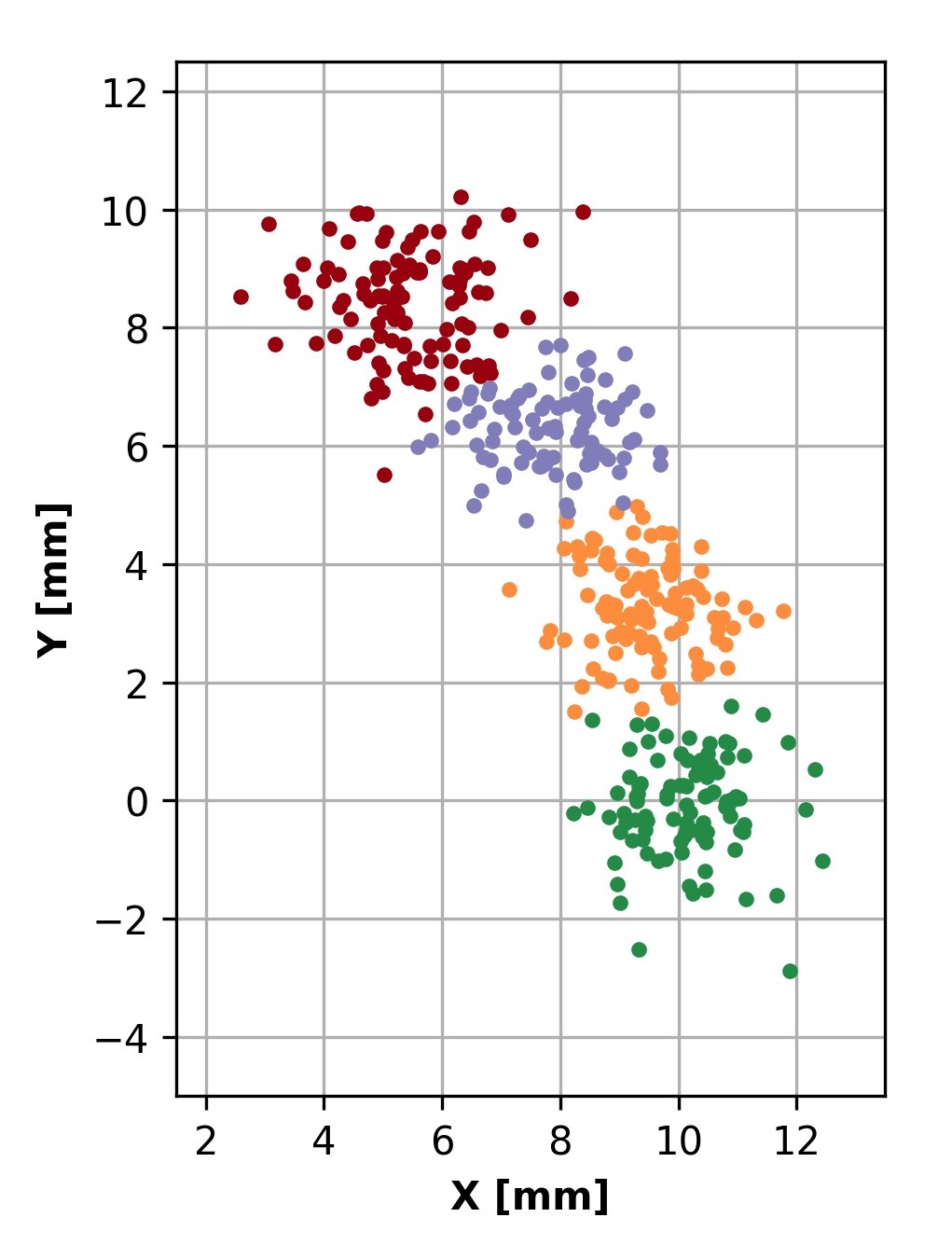}
         \caption{Phase First algorithm}
         \label{fig:PhaseFirst}
     \end{subfigure}\\
     \caption{Examples of clustering results for different algorithms for a simulated data set where spots of ellipticity 1 are separated by $4\sigma$. The clusters' colors serve only to distinguish them from each other. As can be seen, the EM algorithm in Cartesian coordinates (\ref{fig:GMMCart compare}), the Phase First algorithm (\ref{fig:PhaseFirst}), and the Mean Shift algorithm (\ref{fig:MeanShift}) are all able to resolve the 4 spots, whereas the EM algorithm in polar coordinates (\ref{fig:GMMPolar}) and the Variational Bayesian algorithms (\ref{fig:BGMCart}, \ref{fig:BGMPolar}) fail to find all 4 spots. At best, these failures cause further data analysis to be inaccurate, and at worst they prevent further analysis altogether.}
     \label{fig: Compare Algorithms}
\end{figure*}

We found that for single spots rotated $15^{\circ}$ or less relative to the tangent to the ring, all GMMs studied could be used to accurately find spot centers under all tested scenarios, and were more precise by up to a factor of 2.2 than the Mean Shift algorithm. These results are in agreement with the conclusions presented in Ref. \cite{Karthein2021}. The five GMMs studied were also effective for spots separated by at least $5\sigma$. For smaller spot separations, using the Variational Bayesian algorithm of parameter estimation and the EM algorithm with polar coordinates failed to fit to all the spot centers for most scenarios (Fig. \ref{fig: Compare Algorithms}). The EM algorithm in Cartesian coordinates and the Phase First algorithm were able to resolve spots with a separation of $4\sigma$ for all scenarios if there were at most 4 spots. All GMMs became less precise for any amount of noise, but we found that the EM algorithm and Phase First method still performed well for noise intensities less than or equal to one. 

\section{GMM Evaluation with experimental PI-ICR Data}
\label{Real Data}
After testing GMMs using simulated data, we tested their performance on experimental data by performing a consistency check using previously-conducted PI-ICR mass measurements at the CPT mass spectrometer. We checked the masses of $^{163}\text{Gd}$ and $^{162}\text{Tb}$ and their observed isomeric states from Ref. \cite{OrfordNIMB2020} and \cite{OrfordPRC2020}, respectively, using the GMM methods outlined in Section \ref{background}.  These data sets were chosen because the files in each show multiple, non-spherical spots, and some files in the $^{163}\text{Gd}$ data sets show some degree of overlap. They are therefore characteristic representations of typical data sets observed in PI-ICR. The analysis pathway followed the methods as outlined in Ref. \cite{OrfordNIMB2020} and \cite{OrfordPRC2020}, except that instead of using the Mean Shift algorithm and two one-dimensional Gaussian fits to cluster the data and determine spot centers and uncertainties, we used GMMs.

\begin{figure*}[ht!]
     \centering
     \begin{subfigure}[t]{0.45\textwidth}
         \centering
         \includegraphics[width=\textwidth]{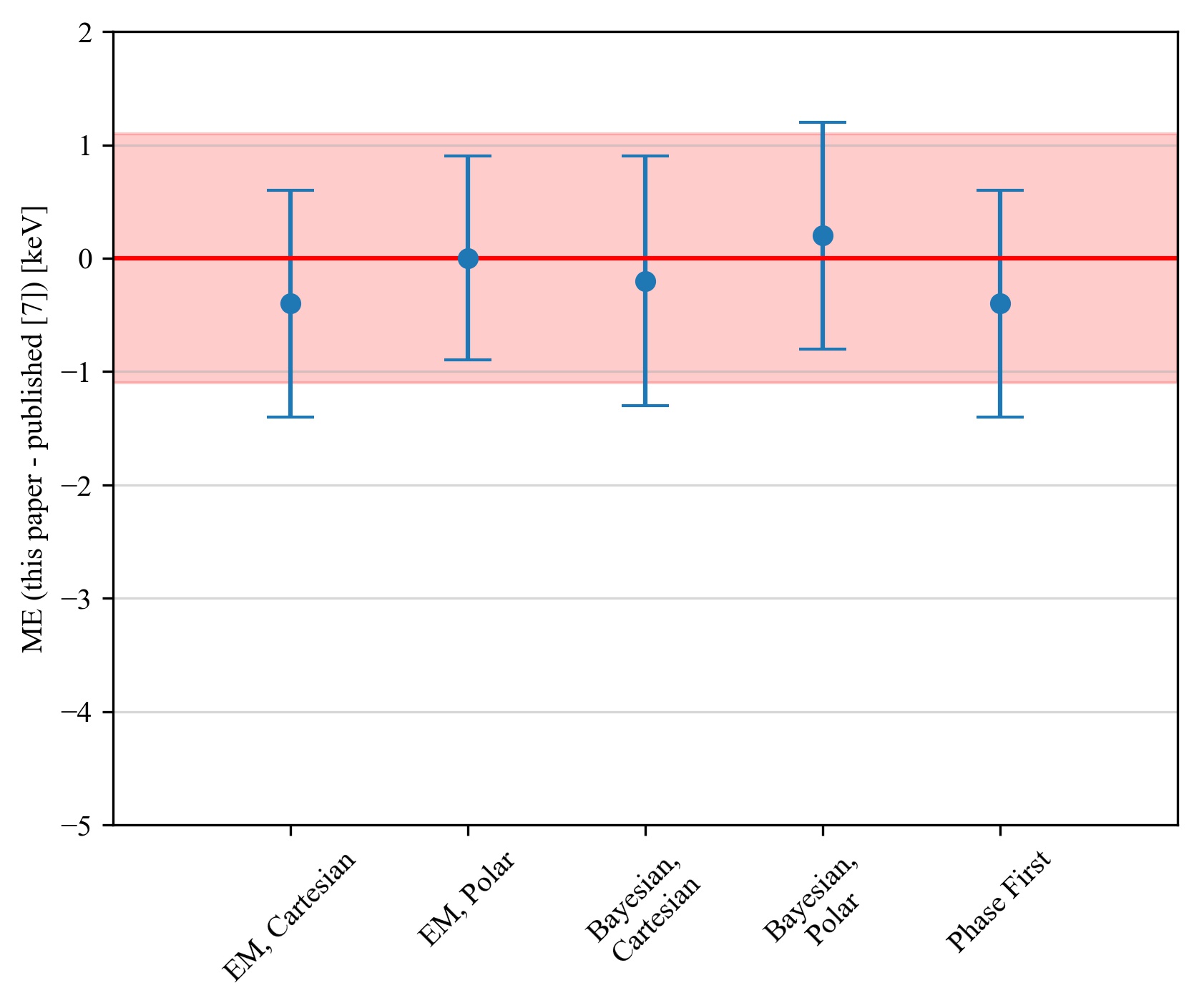}
         \caption{$^{163}\text{Gd}$}
         \label{fig:Gd_g}
         \quad
     \end{subfigure}
     \hfill
     \begin{subfigure}[t]{0.45\textwidth}
         \centering
         \includegraphics[width=\textwidth]{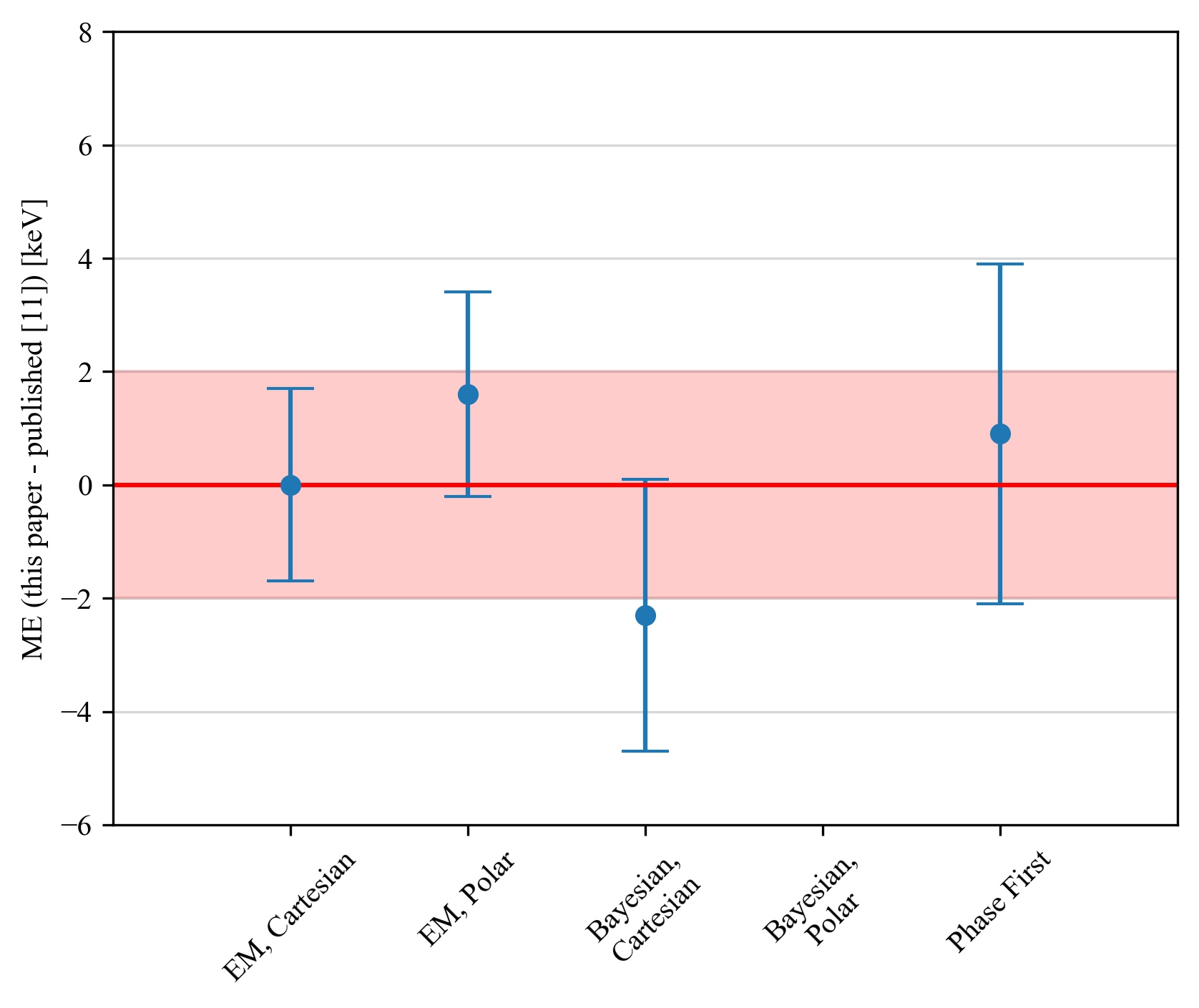}
         \caption{$^{162}\text{Tb}$}
         \label{fig:Tb_g}
     \end{subfigure}\\
     \begin{subfigure}[t]{0.45\textwidth}
         \centering
         \includegraphics[width=\textwidth]{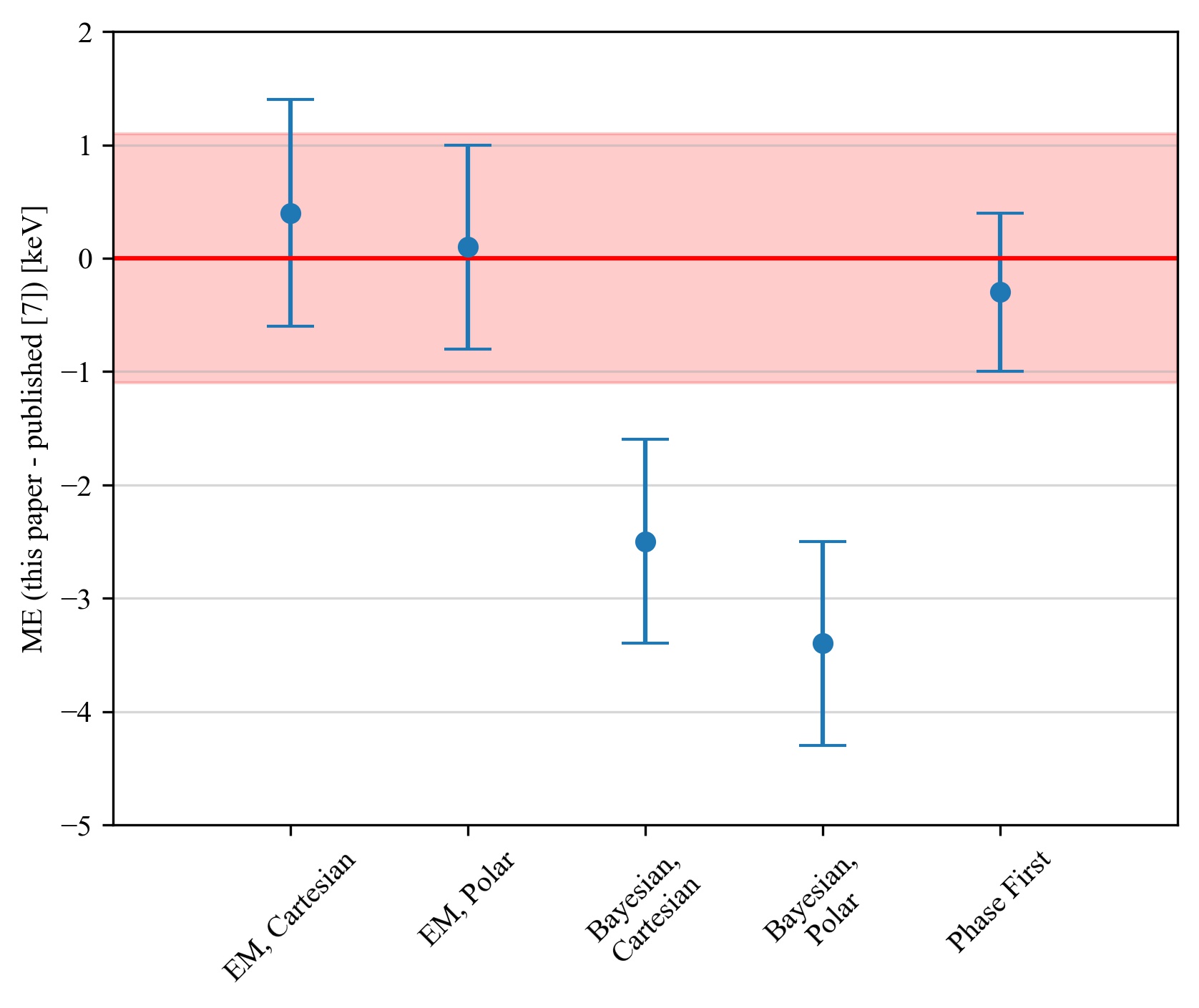}
         \caption{$^{163m}\text{Gd}$}
         \label{fig:Gd_m}
         \quad
     \end{subfigure}
     \hfill
     \begin{subfigure}[t]{0.45\textwidth}
         \centering
         \includegraphics[width=\textwidth]{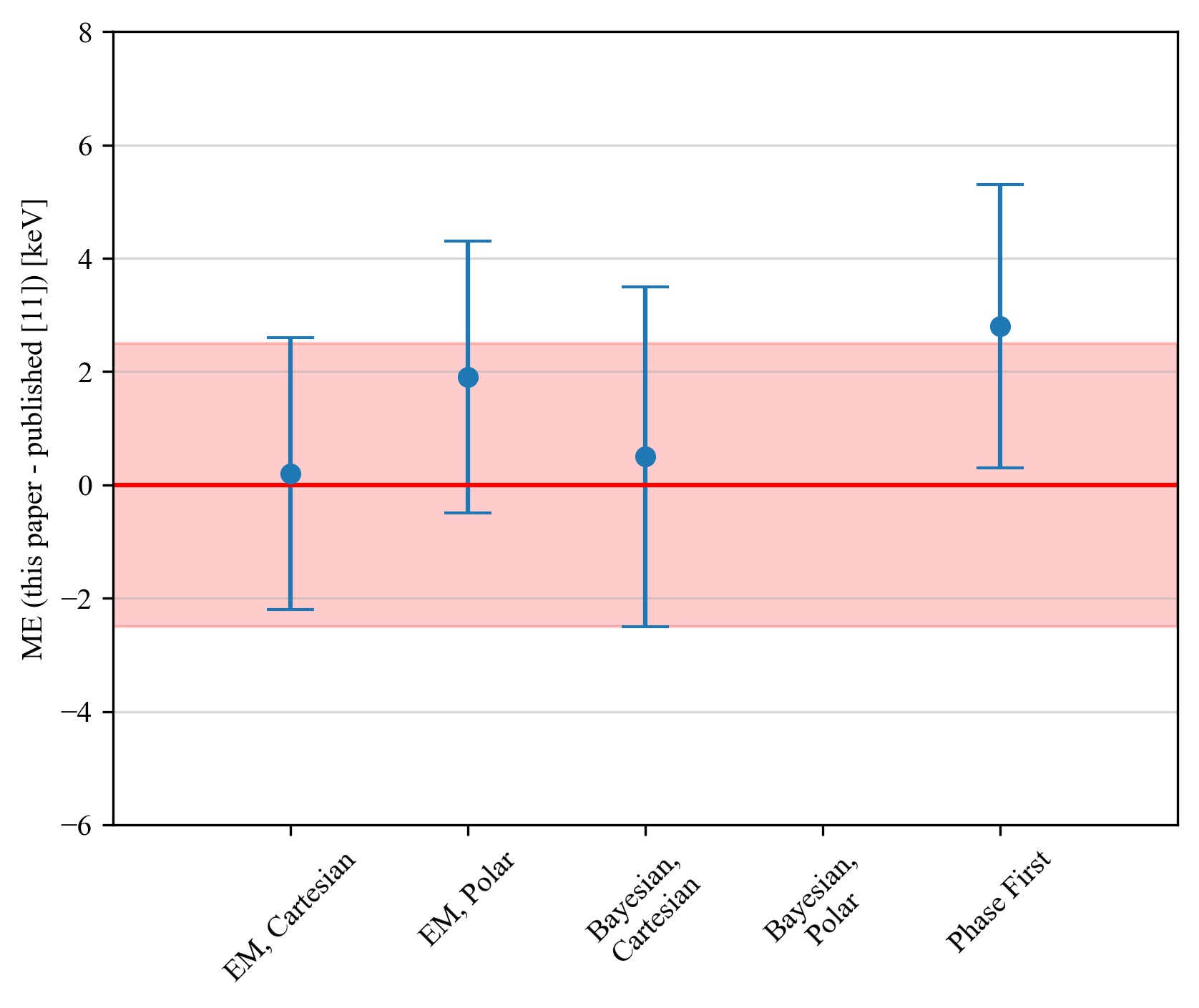}
         \caption{$^{162m}\text{Tb}$}
         \label{fig:Tb_m}
     \end{subfigure}\\
     \caption{The deviation from published results of the mass excesses calculated using GMM algorithms. The red error bands in the $^{163}\text{Gd}$ figures represent results published in Ref. \cite{OrfordNIMB2020}, and the red error bands in the $^{162}\text{Tb}$ figures represent results published in Ref. \cite{OrfordPRC2020}. For $^{162}\text{Tb}$ and $^{162m}\text{Tb}$, there are no results for the Variational Bayesian algorithm in polar coordinates because the algorithm clustered separate spots together in several data files, thus making further analysis impossible.}
     \label{fig:ME compare}
\end{figure*}

The frequency ratio, mass excess, and excitation energy results of this consistency check are presented in Table \ref{tab:Results}. The ``New Result" column shows the new value obtained using the different clustering algorithms, and the ``Deviation" column shows the difference between the new result and the published results. The results for mass excess deviation are also displayed in Fig. \ref{fig:ME compare}. For all nuclides analyzed, the EM parameter estimation method using either Cartesian or Polar coordinates succeeded in reproducing the published results, and when the species of interest was the least populous species in the beam (as in the cases of $^{163m}\text{Gd}$ and $^{162}\text{Tb}$), it also reduced uncertainty by up to a factor of 1.2. 

The Variational Bayesian method using Cartesian coordinates reproduced the published results only when the species of interest was more populous than its counterpart in a different energy state ($^{163}\text{Gd}$ and $^{162m}\text{Tb}$), and in these cases it reproduced the published uncertainty. The Variational Bayesian method using polar coordinates also performed similarly with the $^{163}\text{Gd}$ data. In all other cases, the Variational Bayesian method displayed an emergent property in that it tended to cluster samples together that should have been in separate clusters. Fig. \ref{fig:BGMCart} and \ref{fig:BGMPolar} show behavior similar to this that was observed in the simulated data sets. In short, we observed that the Variational Bayesian method less effective at finding clusters in PI-ICR data sets. The severity of this effect caused the mass excess calculations for $^{163m}\text{Gd}$ using Variational Bayesian methods to be $2-3\sigma$ low, and prevented the Tb clustering results from the Variational Bayesian method using polar coordinates from being analyzed further. We conclude that the EM algorithm is the superior method for fitting GMMs to PI-ICR data.

The Phase First algorithm reproduced the $^{163}\text{Gd}$ results and decreased the uncertainty in mass excess for $^{163m}\text{Gd}$ by a factor of 1.6, but increased uncertainty by a factor of 1.2 when analyzing the $^{162}\text{Tb}$ and $^{162m}\text{Tb}$ data.

\section{Conclusion and Outlook} \label{Outlook}
We have shown that Gaussian mixture models show promise for facilitating PI-ICR data analysis. In both simulated data and experimental data, the EM algorithm for GMM parameter estimation, with input data in Cartesian coordinates, was most effective, with the ability to resolve spots separated by $4\sigma$. This method proved superior to the Mean Shift algorithm in this respect, even though it didn't require assumptions to be made about spot size, density, or quantity. The results from the EM algorithm also do not supersede previously published results because the difference between the two is within $0.4\sigma$, and thus insignificant. The results from Ref. \cite{OrfordNIMB2020} and \cite{OrfordPRC2020} should continue to be used.

The Variational Bayesian method, on the other hand, was shown to be worse at finding the clusters. This effect, while not always observed, caused the algorithm's results to be multiple standard deviations away from previously published results. In the worst case, the new results obtained using the Variational Bayesian method couldn't be further analyzed. The EM algorithm should therefore be the preferred clustering and fitting algorithm with PI-ICR data. If a spectrum has a single spot and no noise, our test results agree with the conclusions reached by the authors of Ref. \cite{Karthein2021}. However, for spectra with multiple spots, non-spherical spots, or noise, the EM algorithm is more effective for clustering and fitting PI-ICR data.

One area for further investigation with GMMs is how they perform when there are spots in the data that appear to have ``tails", because these spots aren't normally distributed with respect to the phase dimension. Preliminary testing on experimental tail spots with GMMs has shown results consistent with the results found using Mean Shift clustering.

Tail spots aside, we conclude that using the EM algorithm with GMMs is the preferred clustering method for use in PI-ICR data analysis given sufficient spot separation, quantity, and noise level. Consequently, we published a Python package called ``piicrgmms" specifically for use in PI-ICR experiments \cite{piicrgmms}. The package includes functions for clustering PI-ICR data using any of the GMMs tested for this paper, and producing images like those shown in Fig. \ref{fig:GMM Algorithm}.

\section{Acknowledgement}
This work was supported by the U.S. Department of Energy, Office of Science, Office of Nuclear Physics under contract No. DE-AC02-06CH11357, by the Natural Sciences and Engineering Research Council (NSERC) of Canada under Contract No. SAPPJ2018-00028, and by the U.S. Department of Energy, Office of Science, Office of Workforce Development for Teachers and Scientists (WDTS) under the Science Undergraduate Laboratory Internships Program (SULI).

\bibliographystyle{model1a-num-names}
\bibliography{Bib}

\begin{landscape}
\begin{table}[t]
\caption{The results for frequency ratio ($\nu_c^{\text{cal}}/\nu_c$), mass excess ME, and isomeric excitation energy $E_x$ calculated from a consistency check of previously published CPT mass spectrometer PI-ICR data using GMMs. The results from the mass excess deviation column are displayed in Fig. \ref{fig:ME compare}.}
\label{tab:Results}
\centering
\begin{threeparttable}
    \begin{tabular}{@{\extracolsep{4pt}}c|c|cccccc@{}}
    \cline{1-8}
        \noalign{\vskip\doublerulesep
        \vskip-\arrayrulewidth} 
    \cline{1-8}
        \noalign{\vskip\doublerulesep
        \vskip-\arrayrulewidth}
    Nuclide & \multirow{2}{*} {Algorithm} & \multicolumn{2}{c}{$\nu_c^{\text{cal}}/\nu_c$} & \multicolumn{2}{c}{ME [keV]} & \multicolumn{2}{c}{$E_x$ [keV]} \\ \cline{3-4} \cline{5-6} \cline{7-8}
    (Calibrant) & & New Result & Deviation ($\times 10^{-9}$) & New Result & Deviation & New Result & Deviation \\ \hline
    \multirow{4}{*}{$^{163}\text{Gd}^{2+}$\tnotex{tn:NIMB}}
    & EM, Cartesian & 0.994 549 904 2 (61) & -2.7 (94) & -61 389.5 (10) & -0.4 (15) & & \\
    \multirow{4}{*}{($^{82}\text{Kr}^{+}$)}
    & EM, Polar & 0.994 549 900 6 (60) & -6.3 (94) & -61 389.1 (9)\phantom{1} & \phantom{-}0.0 (14) & & \\
    & Bayesian, Cartesian & 0.994 549 905 5 (68) & -1.4 (99) & -61 389.3 (11) & -0.2 (16) & & \\
    & Bayesian, Polar & 0.994 549 908 4 (63) & \phantom{-}1.5 (96) & -61 388.9 (10) & \phantom{-}0.2 (15) & & \\
    & Phase First & 0.994 549 904 2 (61) & -2.7 (94) & -61 389.5 (10) & -0.4 (15) & & \\ \cline{1-8}
    \multirow{4}{*}{$^{163m}\text{Gd}^{2+}$\tnotex{tn:NIMB}} & EM, Cartesian & 0.994 550 821 9 (60) & \phantom{-}2.9 (94) & -61 249.5 (10) & \phantom{-}0.4 (15) & 140.0 (14) & \phantom{-}0.8 (21) \\
    \multirow{4}{*}{($^{82}\text{Kr}^{+}$)}
    & EM, Polar & 0.994 550 819 7 (56) & \phantom{-}0.7 (91) & -61 249.8 (9)\phantom{1} & \phantom{-}0.1 (14) & 139.3 (13) & \phantom{-}0.1 (21) \\
    & Bayesian, Cartesian & 0.994 550 802 6 (60) & -16.4 (94) & -61 252.4 (9)\phantom{1} & -2.5 (14) & 136.9 (14) & -2.3 (21) \\
    & Bayesian, Polar & 0.994 550 797 1 (56) & -21.9 (91) & -61 253.3 (9)\phantom{1} & -3.4 (14) & 135.6 (13) & -3.6 (21) \\
    & Phase First & 0.994 550 816 9 (40) & -2.1 (82) & -61 250.2 (7)\phantom{1} & -0.3 (13) & 139.3 (12) & \phantom{-}0.1 (20) \\ \cline{1-8}
    \multirow{4}{*}{$^{162}\text{Tb}^{2+}$\tnotex{tn:PRC}} & EM, Cartesian & 1.037 384 012 (11) & \phantom{-1}0 (18) & -65 879.4 (17) & \phantom{-}0.0 (26) & & \\
    \multirow{4}{*}{($\text{C}_6\text{H}_6^+$)} & EM, Polar & 1.037 384 023 (12) & \phantom{-}11 (18) & -65 877.8 (18) & \phantom{-}1.6 (27) & & \\
    & Bayesian, Cartesian & 1.037 383 996 (16) & -16 (21) & -65 881.7 (24) & -2.3 (28) & & \\
    & Bayesian, Polar & & & & & & \\
    & Phase First & 1.037 384 018 (20) & \phantom{-1}6 (24) & -65 878.5 (30) & \phantom{-}0.9 (36) & & \\ \cline{1-8}
    \multirow{4}{*}{$^{162m}\text{Tb}^{2+}$\tnotex{tn:PRC}} & EM, Cartesian & 1.037 385 977 (17) & \phantom{-1}2 (24) & -65 593.7 (24) & \phantom{-}0.2 (35) & 285.7 (29) & \phantom{-}0.0 (43) \\
    \multirow{4}{*}{($\text{C}_6\text{H}_6^+$)} & EM, Polar & 1.037 385 989 (16) & \phantom{-}14 (23) & -65 592.0 (24) & \phantom{-}1.9 (35) & 285.8 (30) & \phantom{-}0.3 (44) \\
    & Bayesian, Cartesian & 1.037 385 979 (21) & \phantom{-1}4 (27) & -65 593.4 (30) & \phantom{-}0.5 (39) & 288.3 (38) & \phantom{-}2.8 (50) \\
    & Bayesian, Polar & & & & & & \\
    & Phase First & 1.037 385 995 (17) & \phantom{-}20 (24) & -65 591.1 (25) & \phantom{-}2.8 (35) & 287.4 (39) & \phantom{-}1.9 (50) \\ \cline{1-8}
        \noalign{\vskip\doublerulesep
        \vskip-\arrayrulewidth} \cline{1-8}
    \end{tabular}
    \begin{tablenotes}
    \item[a] \label{tn:NIMB}{Compared to results from Ref. \cite{OrfordNIMB2020}}
    \item[b] \label{tn:PRC}{Compared to results from Ref. \cite{OrfordPRC2020}}
    \end{tablenotes}
\end{threeparttable}
\end{table}
\end{landscape}

\end{document}